\documentclass[twocolumn]{aastex631}
\usepackage{amsmath}
\usepackage{comment}
\usepackage{booktabs}
\usepackage[caption=false]{subfig}
\usepackage{savesym}
\savesymbol{tablenum}
\usepackage{siunitx}
\restoresymbol{SIX}{tablenum}

\usepackage{natbib}
\usepackage{verbatim}
\usepackage{mathptmx}  
\usepackage{url}
\usepackage{numprint}
\usepackage{ulem}

\newcommand{\lpicola}{\rm L{\scriptsize -PICOLA}}
\newcommand{\gadget}{\rm G{\scriptsize ADGET}}
\newcommand{\mdpatchy}{MD-{\scriptsize PATCHY}}

\newcommand{\shmr}{\texttt{Fixed SHMR}}
\newcommand{\sham}{\texttt{SHAM}}

\definecolor{mycol}{rgb}{0.05, 0.4, 0.1}
\def\my{}

\received{July XX, 2024}
\revised{July XX, 2024}
\accepted{August XX, 2024}

\begin{document}

\title{Inferring Cosmological Parameters on SDSS via Domain-Generalized Neural Networks and Lightcone Simulations}

\correspondingauthor{Ji-hoon Kim}
\email{mornkr@snu.ac.kr}

\author[0009-0006-4981-0604]{Jun-Young Lee}
\affiliation{Institute for Data Innovation in Science, Seoul National University, Seoul 08826, Korea}
\affiliation{Center for Theoretical Physics, Department of Physics and Astronomy, Seoul National University, Seoul 08826, Korea}

\author[0000-0003-4464-1160]{Ji-hoon Kim} 
\affiliation{Institute for Data Innovation in Science, Seoul National University, Seoul 08826, Korea}
\affiliation{Center for Theoretical Physics, Department of Physics and Astronomy, Seoul National University, Seoul 08826, Korea}
\affiliation{Seoul National University Astronomy Research Center, Seoul 08826, Korea}

\author[0000-0002-9144-1383]{Minyong Jung}
\affiliation{Center for Theoretical Physics, Department of Physics and Astronomy, Seoul National University, Seoul 08826, Korea}

\author[0000-0003-4597-6739]{Boon Kiat Oh}
\affiliation{Department of Physics, University of Connecticut, Storrs, CT 06269, USA}

\author[0000-0003-3977-1761]{Yongseok Jo}
\affiliation{Columbia Astrophysics Laboratory, Columbia University, New York, NY 10027, USA}
\affiliation{Center for Computational Astrophysics, Flatiron Institute, New York, NY 10010, USA}

\author{Songyoun Park}
\affiliation{Center for Theoretical Physics, Department of Physics and Astronomy, Seoul National University, Seoul 08826, Korea}

\author[0000-0002-6810-1778]{Jaehyun Lee}
\affiliation{Korea Astronomy and Space Science Institute,  Daejeon 34055, Republic of Korea}

\author[0000-0001-5082-9536]{Yuan-Sen Ting}
\affiliation{Research School of Astronomy \& Astrophysics, Australian National University, Canberra, ACT 2611, Australia}
\affiliation{School of Computing, Australian National University, Acton ACT 2601, Australia}
\affiliation{Department of Astronomy, The Ohio State University, Columbus, OH 43210, USA}
\affiliation{Center for Cosmology and AstroParticle Physics (CCAPP), The Ohio State University, Columbus, OH 43210, USA}

\author[0000-0003-3428-7612]{Ho Seong Hwang}
\affiliation{Seoul National University Astronomy Research Center, Seoul 08826, Korea}
\affiliation{Astronomy Program, Department of Physics and Astronomy, Seoul National University, Seoul 08826, Korea}



\begin{abstract}
We present a proof-of-concept simulation-based inference on $\Omega_{\rm m}$ and $\sigma_{8}$ from the SDSS BOSS LOWZ NGC catalog using neural networks and domain generalization techniques without the need of summary statistics. Using rapid lightcone simulations, \lpicola{}, mock galaxy catalogs are produced that fully incorporate the observational effects. The collection of galaxies is fed as input to a point cloud-based network, \texttt{\texttt{Minkowski-PointNet}}. We also add relatively more accurate \gadget{} mocks to obtain robust and generalizable neural networks. By explicitly learning the representations which reduces the discrepancies between the two different datasets via the semantic alignment loss term, we show that the latent space configuration aligns into a single plane in which the two cosmological parameters form clear axes.
Consequently, during inference, the SDSS BOSS LOWZ NGC catalog maps onto the plane, demonstrating effective generalization and improving prediction accuracy compared to non-generalized models.
Results from the ensemble of 25 independently trained machines find $\Omega_{\rm m}{=}0.339{\pm}0.056$ and $\sigma_{8}{=}0.801{\pm}0.061$, inferred only from the distribution of galaxies in the lightcone slices without relying on any indirect summary statistics.  
A single machine that best adapts to the \gadget{} mocks yields a tighter prediction of $\Omega_{\rm m}{=}0.282{\pm}0.014$ and $\sigma_{8}{=}0.786{\pm}0.036$. 
We emphasize that adaptation across multiple domains can enhance the robustness of the neural networks in observational data.
\end{abstract}



\section{Introduction} \label{sec:intro}

Following its success in explaining the clustering of matter over a wide range of scales, the $\Lambda$CDM model has now ushered in the era of precision cosmology.
The small perturbations imprinted in the cosmic microwave background grow as cold dark matter falls into and deepens potential wells. 
Small structures gravitationally evolve to create the characteristic cosmic webs and voids referred to as the large scale structures \citep[LSS;][]{PeeblesLSS, DavisLSS, BondLSS}, which are observable in galaxy surveys \citep{Lapparent1986, Geller1989}.
The LSS serves as a widely used probe for constraining the cosmological parameters constituting the $\Lambda$CDM model, as it maps the distribution and motion of matter throughout the universe over time. 
Over the past few decades, a series of galaxy redshift surveys have been conducted extensively to trace the distribution of galaxies and the growth history of LSS across a large spatial extent and depth \citep{Huchra1983, York2000, Colless2001, HectoMap}. 

Considering the galaxy distribution as a (biased) proxy for the total matter content of the universe, power spectrum multipoles and $n$-point correlation functions ($n$-pCF) can be derived to express matter clustering at different scales.
These summary statistics serve as essential components in the development of mock catalogs and in the inference of cosmological parameters. 
The construction of survey-specific mocks, which mimic similar summary statistics and the geometry of the survey, imposes constraints on certain cosmological parameters \citep{Kitaura2016, QPM, Saito2016}. 
Through high-resolution simulations in large volumes and by assigning adequate band magnitudes and spectroscopic information, generic catalogs applicable to various observational surveys can also be generated \citep{MiceI, MiceII, MiceIII, UCHUUSDSS}. 
Other than producing the mocks that best match the observational catalog, derived summary statistics from realizations simulated with varying cosmology can be compared with the observational counterpart to make inference on the cosmological parameters, an approach referred to as simulation-based inference \citep{QUIJOTE, Molino}. While these cited works rely on predefined summary statistics, the simulation-based inference framework allows for the potential use of raw inputs together with the neural networks' flexible featurizations, which permits the exploration beyond summary statistics.

With the advent of artificial intelligence and machine learning, simulation-based inference of cosmological parameters has been accelerated. 
This involves inferring cosmological parameters from simulations by matching summary statistics or features, with neural networks serving as an option alongside more traditional measures of statistical inference such as Markov chain Monte Carlo \citep[MCMC;][]{DELFI, Jeffrey2020}. 
In particular, classic summary statistics such as the $n$-pCF and power spectra, which convey limited information about the matter distribution of the universe, can be replaced with features extracted by neural networks that capture much more complex information engraved inside \citep{Shao2023}.
Attributed to this capability of extracting rich information not hinted at in the summary statistics, simulation-based inference with neural networks has shown the possibility of producing tight predictions on the cosmological parameters \citep{Lesmos2023}. 
Therefore, the importance of simulation-based inference is being recognized as it can serve as an alternative for verifying and possibly resolving tensions in the cosmological parameters predicted from CMB observations and galaxy surveys, especially concerning $H_{0}$ and $S_{8}\equiv \sigma_{8}\sqrt{\Omega_{\rm m}/0.3}$ \citep{Tension_anchoroqui}.

In this context, AI-driven projects have been launched to perform diverse tasks, including parameter estimation \citep{camels_data_release1, CAMELS-ASTRID28, Gigantes}. 
Especially in the estimation of cosmological parameters, 21-cm tomography light cones \citep{Neutsch2022}, weak lensing (WL) convergence and shear maps \citep{Fluri2018, Fluri2019, Fluri2022, Kacprzak2022, Lu2023}, dark matter density fields \citep{Pan2019, lazanu2021, Giri2023, Hortua2023}, and halo catalogs \citep{Ravanbakhsh2016, Mathuriya2019, Ntampaka2020, hwang2023, Shao2023} were utilized as inputs for various neural network architectures, typically in a traditional supervised learning setup.
In contrast to the direct input of mocks, derived summary statistics such as the $n$-pCF, count-in-cell, void probability function, star formation rate density, and stellar mass functions (SMF) were also used as inputs \citep{Boruah2023, Veronesi2023, Hahn2022, Perez2022, Jo2023}. 
In addition, individual galaxy properties \citep{Cosmo1Gal}, galaxy cluster properties \citep{qiu2023}, or snapshots of galaxy catalogs \citep{deSanti2023} were shown to be useful as inputs for neural networks.

    \begin{figure*}[t]
    \vspace{-3mm}
    \centering
    \includegraphics[width=0.98\textwidth]{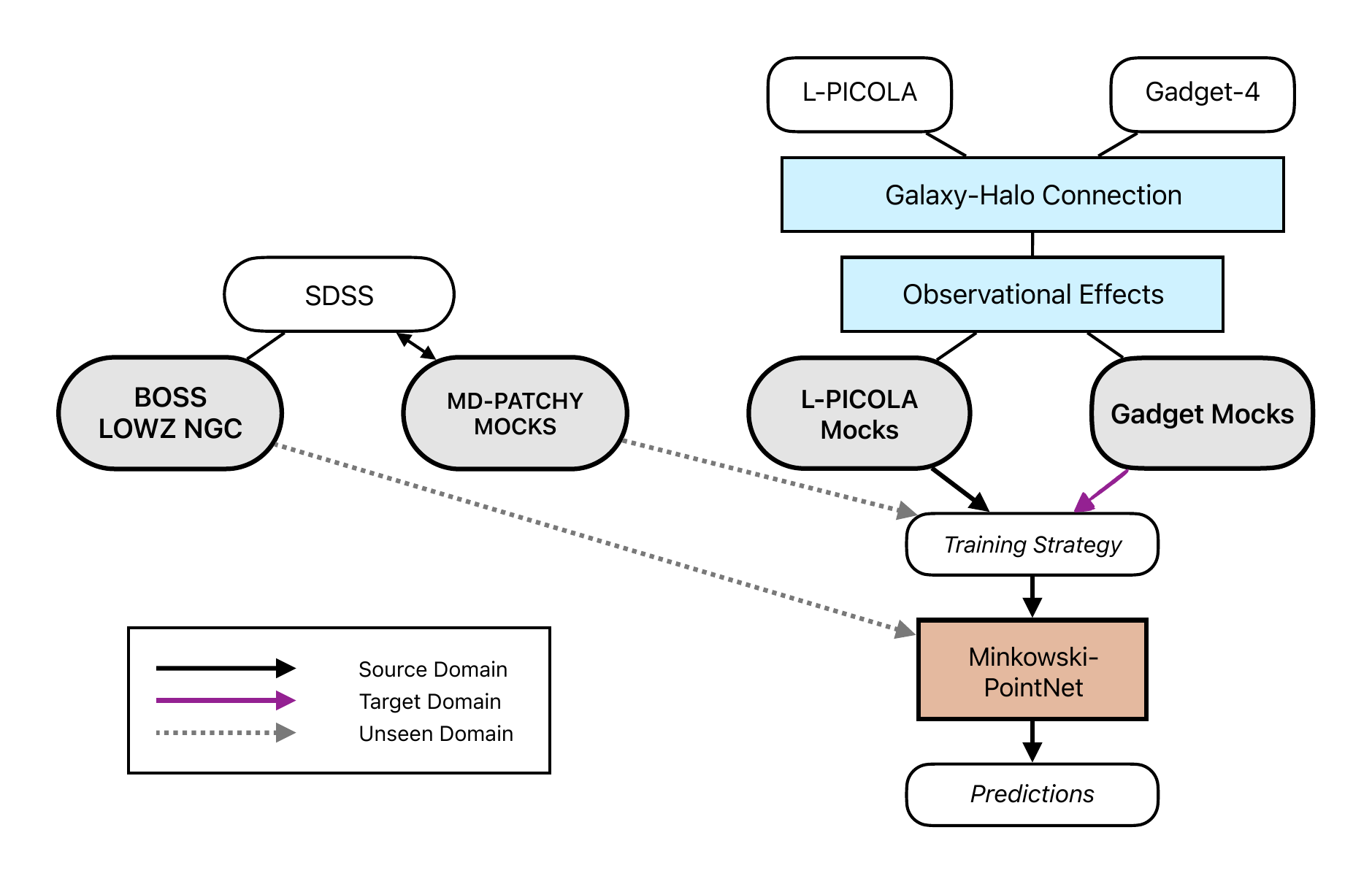}
    \vspace{-5mm}    
    \caption{\label{fig:Overview} Diagram that exhibits the overall structure of this study with simulation and deep learning pipelines. 
    We aim to infer cosmological parameters from the observation-driven catalog, SDSS BOSS LOWZ NGC. 
    We produce two lightcone mock suites, \lpicola{} and \gadget{} mocks, combining $N$-body simulations (sections \ref{sec:colamocks} and \ref{sec:gadget}) and galaxy-halo connection models (Section \ref{sec:GHconnection}) while fully accounting for observational effects (Section \ref{sec:obsEffect}). 
    We then utilize a point-cloud-based network, \texttt{Minkowski-PointNet}, which takes individual galaxies as inputs to predict $\Omega_{\rm m}$ and $\sigma_{8}$, and their errors (Section \ref{sec:neuralnets}). 
    The \lpicola{} mocks (source domain) are trained together with the \gadget{} mocks (target domain) using the training strategy for the domain adaptation and generalization techniques (see Section \ref{sec:SA_strategy}). In this process, we use training strategies to align the representation of each mock (Section \ref{sec:strategies}). The adapted machines are then applied to unseen domains including the fine-tuned \mdpatchy{} mocks and the SDSS BOSS LOWZ NGC sample. The main results including the predictions for the actual observation are shown in Section \ref{sec:pred}.}
    \vspace{2mm}    
    \end{figure*}

Among the listed works, most tested their pipeline on simulated data sets, and only a few successfully generalized their neural networks to the actual observational data. 
\citet{Hahn2022} and \cite{Hahn2023} created a mask autoregressive flow using the power spectrum and bispectrum as summary statistics to provide constraints on  cosmological parameters, based on the SDSS BOSS CMASS catalog \citep{Reid2016}. 
In contrast, \citet{Veronesi2023} leveraged 2-pCF from lognormal mocks as input to fully connected layers (FCL). \citet{Jo2023} used FCL emulators to perform implicit likelihood inference on observed SMF \citep{Leja2020} and SFRD \citep{Leja2022}. 
Parameter inferences using WL convergence maps as probes, including the Kilo Degree Survey \citep{Kids-450,Kids-1000} and Subaru Hyper Suprime-Cam first-year surveys \citep[HSC-Y1;][]{HSC-Y1} were also performed with Convolutional Neural Networks (CNNs) or Graph CNNs \citep[GCNNs;][]{Fluri2019, Fluri2022, Lu2023}. 
Notably, recent studies regard neural networks' outputs of predicted parameters as summary statistics due to their centrally biased nature \citep{Gupta2018, Ribili2019, Fluri2019, Lesmos2023}, and perform additional Bayesian inferences.

In line with efforts to use deep learning for constraining cosmological parameters, this paper aims to perform a proof-of-concept test of conducting cosmological inference using the galaxy redshift survey, {\it without} relying on any indirect summary statistics, but rather utilizing the total raw distribution of galaxies as input to the neural network. 
For this test, we focus mainly on $\Omega_{\rm m}$ and $\sigma_{8}$, which are directly related to the $S_{8}\equiv \sigma_{8}\sqrt{\Omega_{\rm m}/0.3}$ tension as mentioned above. As mentioned in \citet{Hahn2022}, this choice is due to the fact that $\Omega_{\rm m}$ and $\sigma_{8}$ are the parameters that are sensitive to the cosmological information of the clustering galaxies, while others are less constrained.
In order to reduce any artificial priors arising from survey-specific observational biases, we rapidly generate a large mock suite that fully includes observational effects such as redshift space distortion, survey footprint, stellar mass incompleteness, radial selection, and fiber collision in the SDSS BOSS LOWZ Northern Galactic Cap (NGC) catalog. 
Then, using the position and mass information of individual and neighboring galaxies, we make inference on $\Omega_{\rm m}$ and $\sigma_{8}$, again {\it without} relying on any indirect summary statistics.

The biggest difficulty in using the whole galaxy catalog as input instead of the summary statistics is that the selection of codes begets overall differences in the resultant realizations. 
The differences are easily discernible and distinguishable by complex neural networks. 
Consequently, naively merging the different sets of mocks or domains limits the machines to merely learning fragmented domain-specific knowledge. 
Recent studies have tried to address such issues, as machines failing to attain robustness exhibit poor performances and lack predictability on unseen domains \citep{CAMELS-ASTRID28, Shao2023, Roncoli2023}. 
Moreover, as simulated catalogs do not perfectly portray the actual universe, such discrepancies may significantly aggravate the performance of machines onto unseen observed data. 
Especially, the rapid generation of mocks trades off with the inaccuracies compared to the relatively time-consuming simulations, leading to a clear deviation. 
In order to make effective inferences on different types of simulation or domains, the neural network must achieve generalizability. 
This study focuses mainly on extracting and learning unified representations originating from distinct domains and exploiting generalized and integrated knowledge on the observational data.

This paper is organized as follows. 
In Section \ref{sec:sims}, we illustrate the creation of our mock data that thoroughly integrate the observational effects. We produce two suites of mocks using two distinct simulations, \lpicola{} and \gadget{}. The footprint and lightcone slices are shown together with the observational target, the SDSS BOSS LOWZ NGC catalog, and its specific set of mock catalogs, \mdpatchy{}, for comparison.
In Section \ref{sec:neuralnets}, input features and the neural network architecture are introduced together with the training strategies in Section \ref{sec:strategies}, to align the latent space representations of different mocks and achieve domain generalization or robustness. 
In Section \ref{sec:pred}, implicit likelihood estimates in $\Omega_{\rm m}$ and $\sigma_{8}$ using the SDSS BOSS LOWZ NGC catalog are shown. 
We also discuss the impact of fine-tuned \mdpatchy{} mocks on the predictability and generalizability of the machine. 
Finally, the results and the following conclusions are summarized in Section \ref{sec:conc}. 
The overall approach taken by the paper is schematically shown in Figure \ref{fig:Overview}.

\section{\label{sec:sims}GALAXY CATALOG: OBSERVATION AND SIMULATION}

\subsection{The Reference SDSS Catalog}

In this study, we utilize the Baryon Oscillation Spectroscopic Survey \citep[BOSS; ][]{BOSS}, part of SDSS-III \citep{SDSS3}, which extends the previously studied distribution of luminous red galaxies \citep[LRGs; ][]{LRG} from SDSS I/II, adding fainter galaxies and thus larger number densities, for the purpose of measuring baryon acoustic oscillations. The survey consists of the LOWZ \citep{Tojeiro2014} and CMASS \citep{Reid2016} catalogs, which have different color and magnitude cuts. The LOWZ catalog targets galaxies at a low redshift of $z\lesssim 0.4$, while CMASS targets a higher redshift range of $0.4\lesssim z \lesssim 0.7$. 
The LOWZ samples are roughly considered as volume-limited, whereas the CMASS samples, representing `constant mass', are considered volume-limited within the mass and redshift ranges of $M_{\star} > 10^{11.3} \mathrm{M}_{\odot}$ and $z \lesssim 0.6$ \citep{Reid2016, Maraston2013}. Using the {\scriptsize MKSAMPLE} code, the LSS catalogs for both LOWZ and CMASS were created for BOSS DR12, fully equipped with survey masks and random samples. These samples include completeness and weights calculated for the analysis of large-scale structure \citep{Reid2016}.

To account for the stellar mass incompleteness of the survey and to incorporate cosmological information from the stellar masses of galaxies later on, we obtain stellar mass data from the value-added Portsmouth SED-fits catalog \citep{Maraston2013}, assuming a passive evolution model with the Kroupa IMF \citep{Kroupa2001}. Since the Portsmouth SED-fits catalog includes both BOSS and LEGACY targets, we need to select those that are included in the LSS catalog. Following \citet{Rodriguez2016}, we match galaxies using the unique combination of tags {\scriptsize MJD}, {\scriptsize PLATEID}, and {\scriptsize FIBERID}, and then assign the stellar masses from the matched galaxies in the Portsmouth catalog to the corresponding entries in the LSS catalog.

In this work, we use the Northern Galactic Cap (NGC) of the LOWZ samples with RA=150$^\circ$--240$^\circ$ and DEC$>$0$^\circ$. The selection of the LOWZ samples and the cropped regions is due to the limited volume of the lightcone simulations that will be used to generate mocks. Using this catalog as a benchmark, we generate mocks that incorporate the same observational effects: redshift space distortions, survey footprint geometry, stellar mass incompleteness, radial selection matching, and fiber collision (see Section \ref{sec:obsEffect} for more information).

\subsection{\label{sec:colamocks}Rapidly Generated Lightcone Mocks, \texorpdfstring{\sc L-\MakeLowercase{picola}}{L-picola}}

\lpicola{} is a rapid dark matter simulation that employs the COmoving Lagrangian Acceleration method \citep[{\rm C{\scriptsize OLA}}; ][]{COLA} and supports on-the-fly generation of lightcones. At the expense of minute errors---2\% in the power spectrum and 5\% in the bispectrum---the code allows for the rapid generation of dark matter distributions in large box sizes \citep{L-PICOLA}. Numerous studies have leveraged on this computational efficiency to produce a vast amount of mock catalogs aimed for diverse observations \citep{HowlettMGS, Howlett2022, Ishikawa2023}. 

In a box volume of (1.2$h^{-1}$Gpc)$^3$ we simulate the evolution of 1200$^3$ dark matter particles on 1200$^3$  meshes. Each particle has a mass of approximately $M_{p}\approx8.3\times 10^{10}\left( \frac{\Omega{\rm m}}{0.3} \right) h^{-1}{\rm M}_{\odot}$. The simulation starts with a 2LPT initial condition generated with {\rm 2LPT{\scriptsize IC}} \citep{2LPTScoccimarro2012} at $z_{\rm{initial}}{=}$9 and progresses in 10 steps to $z{=}0.45$, as \citet{L-PICOLA} suggests for sufficient precision in the resolution adopted here, with 10 lightcone slices generated from $z{=}0.45$ to $z{=}0$. A total of 1500 simulations are produced, incorporating cosmic variance across varying $\Omega_{\rm m}$ and $\sigma_{8}$. Each of the two parameters is randomly sampled from a uniform distribution of $\Omega_{\rm m}{\in}[0.1, 0.5]$ and $\sigma_{8}{\in}[0.6, 1.0]$. We assume $H_0{=}100 h$ km s$^{-1}$ Mpc$^{-1}$ with $h{=}0.674$, $n_s{=}0.96$ following the results from \cite{Planck2018}. We select a realization from one pair of cosmological parameters most similar to the fiducial cosmology of \mdpatchy{} with $\Omega_{\rm m}=0.3067, \sigma_{8}=0.8238$ and name it \texttt{L-PICOLA fiducial}. We obtain the halos using the {\rm R{\scriptsize OCKSTAR}} halo finder \citep{Rockstar} in lightcone mode, considering a minimum number of 10 particles as a seed halo (most detailed layer of subgroup hierarchy determined by the friends-of-friends algorithm). Thus, we impose a cut in the halo mass of $\log(M_{\rm h}/h^{-1}{\rm M}_{\odot}){=}11.45$. Subsequently, the 1500 catalogs are rotated and reflected in six directions following \citet{Ravanbakhsh2016}, generating a total of 9000 realizations referred to as \lpicola{} mocks. These mocks will be further cropped and masked separately according to the observational effects. From this we establish a one-to-one correspondence between the subhalos and galaxies. 

\subsection{\label{sec:gadget}Adaptation: Gravitational $N$-body Simulation Mocks, \texorpdfstring{\sc G\MakeLowercase{adget}}{Gadget}}

The \lpicola{} mocks described in Section \ref{sec:colamocks} lack accuracy in the clustering statistics on small scales compared to full $N$-body simulations (see Section \ref{sec:domain_shift}). Therefore, we similarly generate mocks using \gadget{}-4 \citep{GADGET-4} in lightcone mode, which we refer to as \gadget{} mocks. Although they require more computational time and resources to generate than \lpicola{} mocks, \gadget{} mocks, are generally considered to offer higher fidelity at smaller scales \citep[see][]{L-PICOLA}. Consequently, we use \gadget{} mocks as adaptation standards of the neural networks, to refine the code-specific knowledge from \lpicola{} mocks, implementing a training strategy that aligns the neural networks' extracted representations. For additional details, refer to Section \ref{sec:strategies}.

The simulation resolution is the same as that of mock suites generated with \lpicola{}: a box volume of (1.2$h^{-1}$Gpc)$^3$ and 1200$^3$ dark matter particles with a softening length of 10$h^{-1}$kpc. The simulation initiates with a 2LPT initial condition generated with {\rm N-{\scriptsize GenIC}} \citep{NGenIc} at z$_{initial}{=}$10, similar to \lpicola{}, and ends at $z{=}0$.\footnote{We acknowledge that starting a full $N$-body simulation, \gadget{}, at low redshifts may lead to inaccuracies, unlike \lpicola{}, despite the reduction of computational resources. The choice of the initial redshift was based on the comparative analyses presented in \cite{L-PICOLA}. We leave such improvements to be addressed in our future work.} The cosmological parameters of the fiducial run, \texttt{GADGET fiducial}, are set to be identical to those of \mdpatchy{} mocks in Section \ref{sec:mdpatchy}: $\Omega_{\rm m}{=}0.307115$, $\sigma_{8}{=}0.8288$, and $h{=}0.6777$, with other parameters fixed to the previously stated values. Furthermore, in order to test the machine's predictability for non-fiducial mocks, we produce \texttt{GADGET low} with $\Omega_{\rm m}{=}0.2$, $\sigma_{8}{=}0.7$ and \texttt{GADGET high} with $\Omega_{\rm m}{=}0.4$, $\sigma_{8}{=}0.8$. We generate 6 samples each by rotating and reflecting the three \gadget{} simulations, totalling 18 samples.

    \begin{figure}[t]
    \centering
    \includegraphics[width=0.46\textwidth]{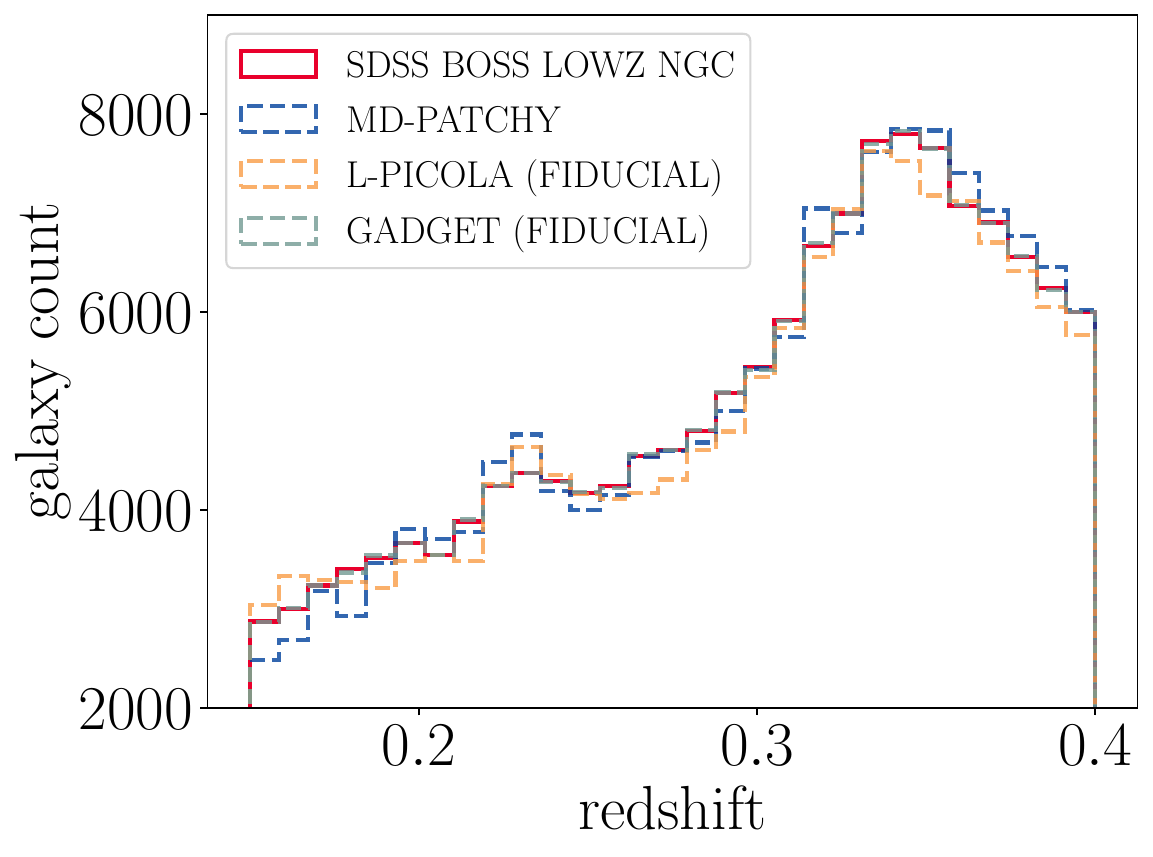}
    \caption{\label{fig:RadialSelection}Count of galaxies per radial bins for SDSS BOSS LOWZ NGC (\textit{red solid line}), \texttt{L-PICOLA fiducial} (\textit{orange dashed line}), \texttt{GADGET fiducial} (\textit{green dashed line}), and averaged count for all 2048 \mdpatchy{} (\textit{blue dashed line}). 
    The radial bins from redshift 0.15 to 0.4 are defined to evenly divide the redshift space volume. 
    All lightcone mocks with fiducial cosmological parameters exhibit a consistent number of galaxies across different radial bins compared to the SDSS BOSS LOWZ NGC catalog.  
    See Section \ref{sec:obsEffect} for more information.}
    \end{figure}

    \begin{figure*}[t!]
    \centering
    \includegraphics[width=0.86\textwidth]{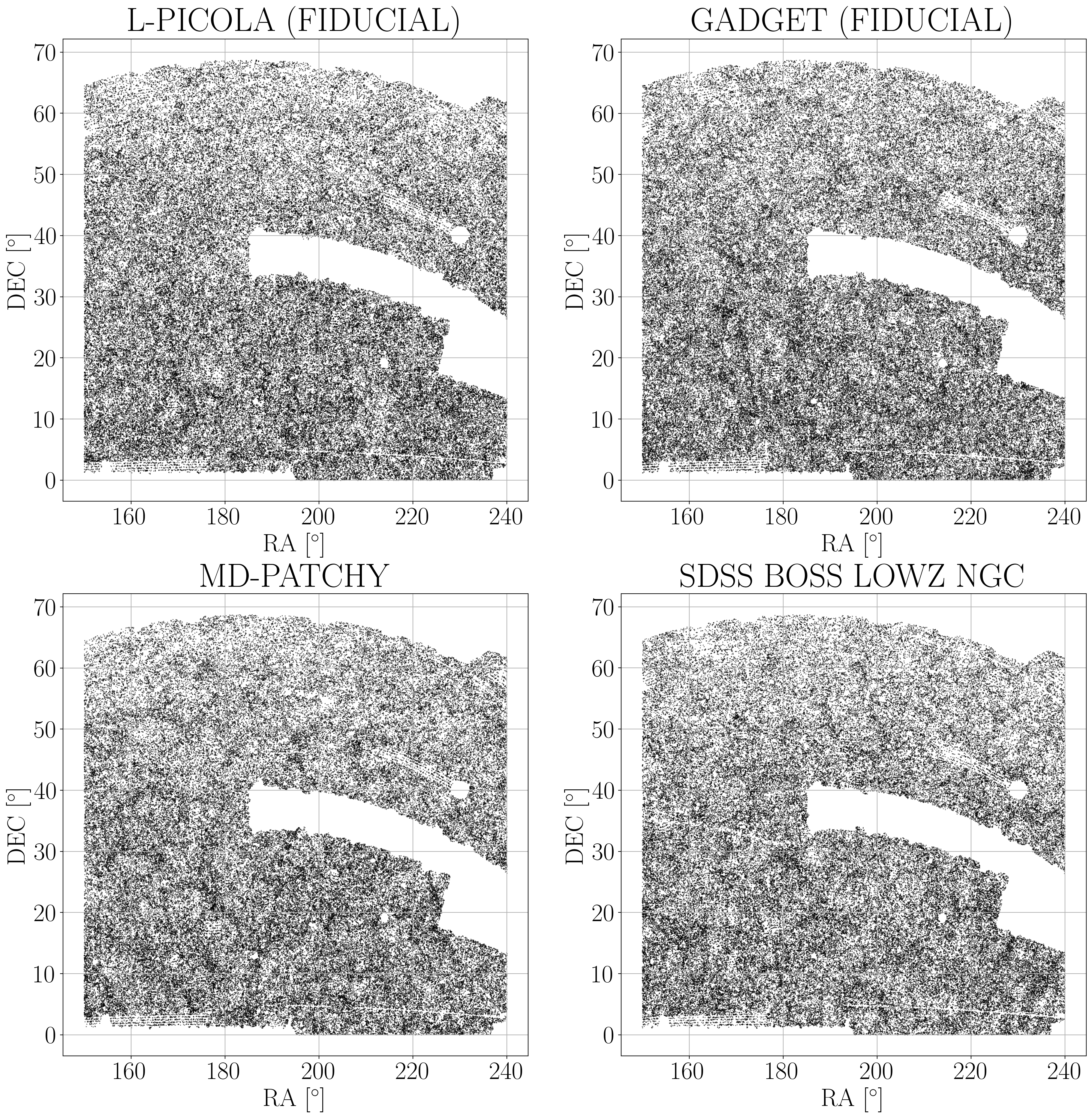}
    \caption{\label{fig:footPrint} The footprints of a single realization from \texttt{L-PICOLA fiducial} (\textit{top left}), \texttt{GADGET fiducial} (\textit{top right}), \mdpatchy{} (\textit{bottom left}), and SDSS BOSS LOWZ NGC catalog (\textit{bottom right}). 
    The same acceptance and veto masks are employed to reproduce the overall topology. 
    We further cut the region into RA=150$^\circ$--240$^\circ$ and DEC$>$0$^\circ$. 
    See Section \ref{sec:obsEffect} for more information.}
    \vspace{2mm}
    \end{figure*}

    \begin{figure*}[t!]
    \centering
    \includegraphics[width=0.86\textwidth]{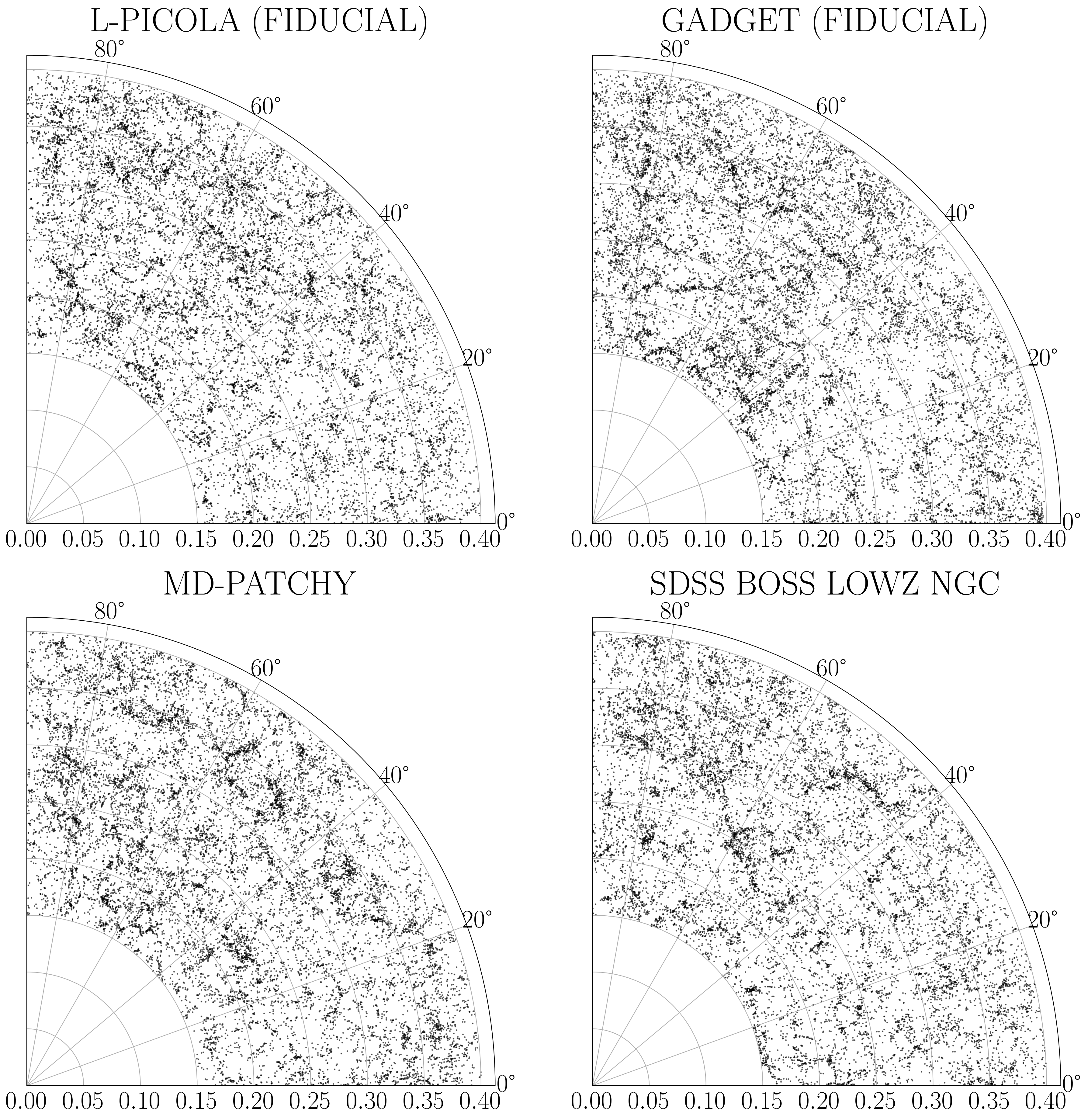}
    \caption{\label{fig:slice} Lightcone slices of Figure \ref{fig:footPrint} from 0$^\circ$$<$DEC$<$6$^\circ$. 
    See Section \ref{sec:obsEffect} for more information.}
    \vspace{4mm}
    \end{figure*}

\subsection{Adaptation: Fine-Tuned Mocks, \texorpdfstring{\sc MD-\MakeLowercase{patchy}}{MD-patchy} \label{sec:mdpatchy}}

\rm M{\scriptsize ULTI}\rm D{\scriptsize ARK} {\scriptsize PATCHY} Mocks (hereafter \mdpatchy{}) are mock galaxy catalogs designed to match the SDSS-III BOSS survey \citep{Kitaura2016, Rodriguez2016}. They referenced the {\rm B{\scriptsize IG}\rm M{\scriptsize ULTI}\rm D{\scriptsize ARK}} simulation \citep{Kyplin2016}, a $N$-body simulation run on \gadget{}-2 \citep{Gadget-2}. The halos from the {\rm B{\scriptsize IG}\rm M{\scriptsize ULTI}\rm D{\scriptsize ARK}} are populated using the stochastic halo abundance matching technique and the observational effects including redshift space distortion, survey footprint, stellar mass incompleteness, radial selection, and fiber collision are considered using the {\scriptsize SUGAR} code \citep{Rodriguez2016}. The reference catalog is used to calibrate {\scriptsize PATCHY} \citep{Patchy2013}, which employs augmented Lagrangian Perturbation Theory \citep[ALPT; ][]{ALPT} to generate dark matter fields. These fields are biased and the halo masses are identified using the {\scriptsize HADRON} code \citep{Hadron}, which takes the halos' environmental information into account. The halo catalog is further processed into galaxy mocks using the halo abundance matching procedure in the {\scriptsize SUGAR} code. Specifically, the clustering statistics are fitted by fine-tuning a single parameter--the scatter in the HAM procedure ($\sigma_{\rm{HAM}}(V_{\rm{peak}}|M_{\star})$), where $M_{\star}$ represents the stellar mass and $V_{\rm{peak}}$ the peak velocity observed throughout the history of the halo. In total, 10240 \mdpatchy{} mocks that mimic the clustering statistics, stellar mass functions, and observational effects are produced. The cosmological parameters used are $\Omega_{\rm m}{=}0.307115$, $\sigma_{8}{=}0.8288$, and $h{=}0.6777$. In this work, we focus on the 2048 mocks of the Northern Galactic Cap (NGC) of the LOWZ samples. Similarly to the \gadget{} mocks in Section \ref{sec:gadget}, the \mdpatchy{} mocks are used as reference mocks for adaptation of the neural networks during the training phase (see Section \ref{sec:strategies} for more information). 

\subsection{Galaxy-Halo Connection\label{sec:GHconnection}}

The galaxy-halo connection is a crucial statistical relation that summarizes the interplay between gravitational evolution and baryonic physics in galaxies and halos, widely studied in the fields of galaxy formation and cosmology \citep[see][for review]{GalaxyHaloAnnual}. Numerous approaches in modeling are available, including the halo occupation distribution \citep[HOD;][]{HOD_Peacock_2000, HOD_Berlind_2003}, subhalo abundance matching \citep[SHAM;][]{Kravtsov2004, Conroy2006}, and also the combined models such as subhalo clustering and abundance matching \citep[SCAM;][]{SCAM,SCAMPY}. In the following, we introduce the two galaxy-halo connection methods: the fixed stellar-to-halo-mass relation and the SHAM.\footnote{The galaxy-halo connection models introduced here are indeed simplistic. To account for the detailed connection relation, it may be necessary to track the halo assembly history or apply varying population models by introducing few additional parameters. Here, we focus on proof-of-concept objective, rather than investigating deeply into this complex relation. Such limitations are left for future work.}

\subsubsection{Fixed Stellar-to-Halo-Mass Relation\label{sec:fixed_shmr}}
Here, we adopt the minimal model that connects $N$-body simulations to galaxy catalogs. Assuming a one-to-one galaxy-subhalo correspondence as employed in the previous works \cite[e.g.,][]{Kim2008, Hwang2016}, we impose a fixed stellar-to-halo-mass relation (SHMR) across different realizations. In other words, we assume that the star formation efficiency of galaxies in halos is equivalent across different cosmologies, within the redshift range of this study.\footnote{This is a strong assumption made to derive the stellar masses of each subhalo identified from a dark matter-only simulation, and where cosmology-dependent information intervenes. This is due to the impracticality of performing full hydrodynamic simulations of such a spatial and temporal extent across varying cosmologies. Despite introducing weak dependency, we emphasize that this assumption is made for a proof-of-concept test. For a model free of cosmological priors, refer to the \sham{} model in Section \ref{sec:sham}.}

We use the SHMR obtained by \cite{Girelli2020}, which compares the DUSTGRAIN-\textit{pathfinder} simulation \citep{Dustgrain} with the SMF determined in \citep{IlbertSMF} from the Cosmological Evolution Survey (COSMOS) \citep{COSMOS}. The SHMR is analyzed per different redshift bins to account for the temporal variability of the efficiency, parameterized as
    \begin{equation}
    \frac{M_{\star}}{M_{h}}(z){=}2A(z)\left[\left(\frac{M_{\star}}{M_{A}(z)}\right)^{-\beta(z)}{+}\left( \frac{M_{h}}{M_{A}(z)}\right)^{-\gamma(z)}\right]^{-1}
    \end{equation}
where $M_{h}$ is the halo mass and $A(z)$ is the normalization factor at $M_{A}$, at which the double power-law breaks. Since our mock galaxies are selected within $0.15{<}z{<}0.40$, we utilize the SHMR parameters estimated for $0.2{<}z{<}0.5$. The best-fit parameters are $A(z){=}0.0429$, $M_{A}{=}11.87$, $\beta{=}0.99$, and $\gamma{=}0.669$ when scatter of $\sigma_{r}{=}0.2$ dex is introduced. We will use these parameters, including the 0.2 dex scatter, for this work. 

\subsubsection{Subhalo Abundance Matching\label{sec:sham}}

In Section \ref{sec:fixed_shmr}, the \shmr{} establishes cosmological priors as it selects the specific relation of connecting the halo mass properties to the baryonic physics. To tackle this issue, we alternatively utilize the non-parametric version of \sham{}—a well-known basic galaxy-halo connection, as previously discussed, which is also used for constraining cosmological parameters \citep{Simha:2013hy}. The halo catalogs are painted with stellar masses using a monotonic relation between the simulated halo masses and the stellar masses identified from the observed SDSS BOSS LOWZ NGC catalog. Therefore, the difference between mocks with different cosmologies arises from the clustering of the galaxies instead of stellar mass itself as compared to the \shmr{} model.

We acknowledge that the prescription in our \sham{} model is simplistic and may not fully describe the galaxy-halo connection. Numerous studies on SHAM have employed the historical peak mass or circular velocity of the halo \citep{Reddick2013, Behroozi2013}. However, the nature of the on-the-fly generation of lightcones precludes the possibility of utilizing historical information. In order to bypass such limitations, \cite{Ishikawa2023} use snapshots instead of generating lightcones on-the-fly and employs post-processing to generate lightcones. However, since our focus here is on the proof-of-concept test of inferring cosmological parameters without summary statistics and using neural networks, we accept the inherent crudeness in the galaxy-halo connection model.

    \begin{figure*}[t]
    \centering
    \includegraphics[width=\textwidth]{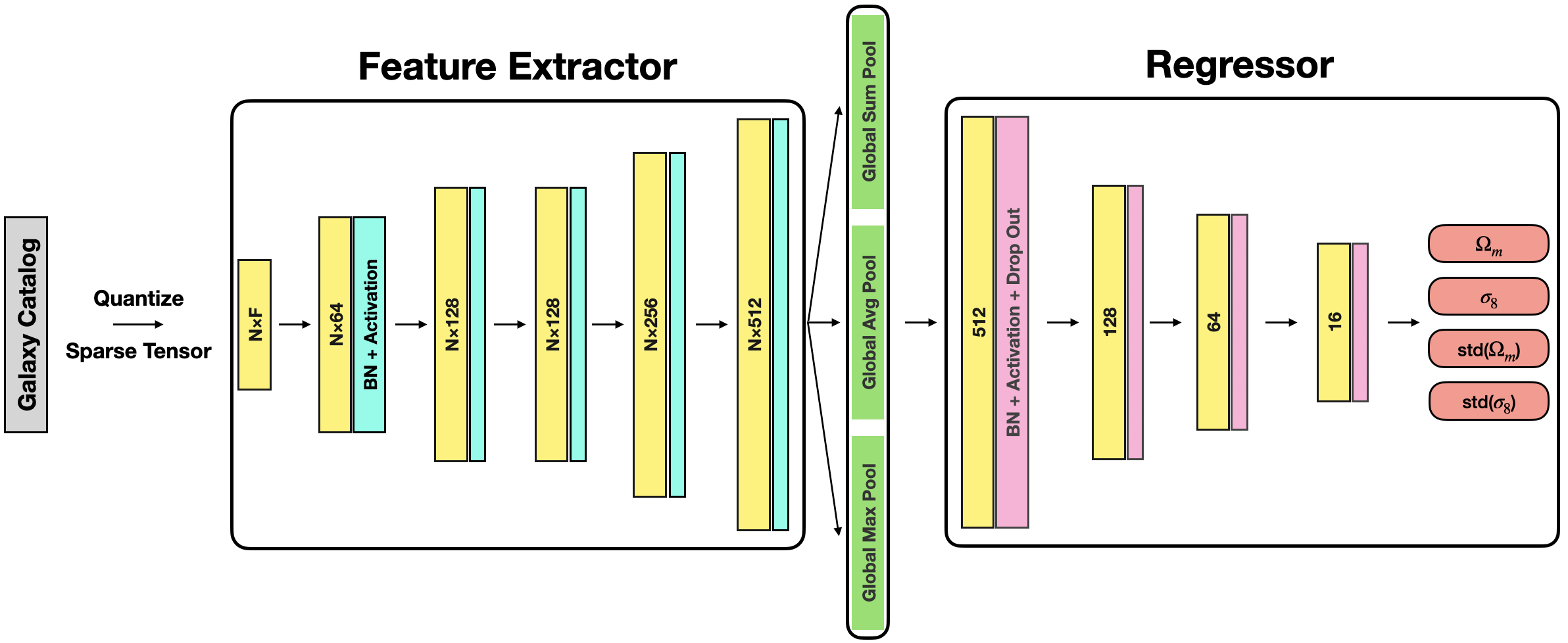}
    \caption{\label{fig:NetworkLayout}Architecture of the \texttt{Minkowski-PointNet} in this work. Each input is transformed into a sparse tensor with $N$ galaxies, each with features $F$, and passed through five consecutive linear layers for feature extraction. Global sum, average and max pooling are done to extract a 1536-dimensional feature vector. Then it passes through the regressor consisting of four linear and dropout layers to predict $\Omega_{\rm m}$, $\sigma_{8}$ and their standard deviations.
    See Section \ref{sec:backbone} for more information.}
    \vspace{5mm}
    \end{figure*}
    
\subsection{Observational Effects\label{sec:obsEffect}}

We include the following observational effects of the SDSS BOSS LOWZ NGC catalog into the \lpicola{} simulations: redshift space distortions, survey footprint geometry, stellar mass incompleteness, radial selection matching, and fiber collision. By fully accounting for these observational effects, we can assess how observables from realizations endowed with different sets of cosmological parameters would have deviated from the actual observation.

Firstly, the positions of the model galaxies are shifted using their peculiar velocities to account for the redshift space distortion \citep[RSD;][]{RSD}. In order to match the footprint geometry of our mocks to that of the SDSS BOSS LOWZ NGC, we apply acceptance and veto masks. Galaxies are filtered out by applying the {\rm M{\scriptsize ANGLE}} masks \citep{Swanson2008} using the {\rm M{\scriptsize AKE}}\_{\rm S{\scriptsize URVEY}} code \citep{Makesurvey}. Next, for both \shmr{} and \sham{} models, we restrict the area of interest to RA=150$^\circ$--240$^\circ$ and DEC$>$0$^\circ$.\footnote{The trimming of the footprint was necessary to accommodate the generation of lightcones in octants of the sphere. This adjustment results in a slight deviation in the data used compared to earlier studies, such as those of \cite{Ivanov2020} and \cite{Hahn2022}. Nonetheless, we expect these differences to be minimal, given the modest nature of the change.}

For the \shmr{} model, we further apply the incompleteness in the galaxy stellar mass function of the SDSS BOSS LOWZ NGC catalog, a statistical bias due to the observational constraints of the survey. Here, we apply the incompleteness of the LOWZ NGC sample, which is modeled by \cite{Leauthaud2016}, using the Stripe 82 Massive Galaxy Catalog to measure the SMF. The incompleteness function is shown in Equation \ref{eqn:inc}, where $f$, $\sigma$, and $M_{1}$ are free parameters for fitting. We calculate the interpolated incompleteness using the stellar mass and redshift of the galaxies, and decide whether to use or discard a galaxy based on the result.
\begin{eqnarray}\label{eqn:inc}
c=\frac{f}{2}\left[1+\text{erf}\left( \frac{\log{M_{\star}}/M_{1}}{\sigma} \right) \right]
\end{eqnarray}
After identifying the galaxies that are not observable due to stellar mass incompleteness and survey geometry, we randomly downsample the galaxies to match the radial selection. This is achieved by finely dividing the redshift range into 260 radial bins with equal redshift space volume spacing.

For the \sham{} model we perform massive downsampling. Unlike the \shmr{} model, the SHAM model inherently includes stellar mass incompleteness because we use the observed galaxy catalog, which already has inherent incompleteness, as our reference. Also, we perform massive sampling instead of random sampling in order to match the monotonicity of the \sham{} process. Similarly to the \shmr{} model, the sampled galaxies are filtered once more through the fiber collision algorithm, and then finally assigned with the appropriate stellar masses.  

Furthermore, we mimic the fiber collision in the SDSS BOSS LOWZ NGC catalog. The SDSS galaxy spectra were obtained from fibers inserted into perforated plates. Since the fibers have a finite size with a collision radius of $\ang{;;62}$, a portion of fiber-collided galaxies has not been assigned with any fibers. Using {\scriptsize NBODYKIT} \citep{Nbodykit}, we classify the galaxies into two populations: decollided galaxies (D$_1$) and potentially collided galaxies (D$_2$) \citep{Guo2012} using the angular friends-of-friends algorithm as in \citet{Rodriguez2016}. The actual abundance matching of the \sham{} model is performed after accounting for the fiber collisions in order to fully preserve the number of galaxies. However, for the \shmr{} model, the stellar mass incompleteness already includes the incompleteness due to fiber collisions. Nevertheless, this reduction should be applied since fiber collisions are an important systematic biases in the small-scale geometry of the survey. We consider the potential double-counting of fiber collisions within the stellar mass incompleteness to have a negligible impact on our final results.

Figure \ref{fig:RadialSelection} compares the galaxy count per radial bins for SDSS BOSS LOWZ NGC, \mdpatchy{}, \texttt{L-PICOLA fiducial}, and \texttt{GADGET fiducial} mocks generated with the \shmr{} model. The similarity in the distributions verify the consistency across all three mocks and the observational catalog. In realizations with low $\Omega_{\rm m}$ and $\sigma_{8}$ generated with the \shmr{} model, the absolute number of galaxies is relatively small, and thus the total number of galaxies may be less than that of the fiducial cosmology. Such a deficit can provide critical information to inform the neural network that the real universe is unlikely to have such cosmological parameters. However, the mocks produced by the \sham{} model does not have difference in the total number of galaxies, as it directly matches the observed galaxy mass to the halo catalog. 

Finally, for both \shmr{} and \sham{} models, we restrict the area of interest to RA=150$^\circ$--240$^\circ$ and DEC$>$0$^\circ$. The four panels of Figure \ref{fig:footPrint} show the footprint of the \texttt{L-PICOLA fiducial}, \texttt{GADGET fiducial}, \mdpatchy{}, and the SDSS BOSS LOWZ NGC catalog. Notice that the masks are equally applied, showing the same apparent streaks and holes. Figure \ref{fig:slice} shows the lightcone slices from 0$^\circ$$<$DEC$<$6$^\circ$ for each of the four mocks, with the observational effects fully taken into account.

\section{\label{sec:neuralnets}Neural Network Architecture}

\subsection{Backbone: \texorpdfstring{\tt M\MakeLowercase{inkowski}-P\MakeLowercase{oint}N\MakeLowercase{et}}{Minkowski-PointNet} \label{sec:backbone}}

A large portion of the universe is empty, as galaxies are predominantly clustered along the filaments of the LSS. Therefore, depositing galaxies into uniform voxels can be highly inefficient, resulting in many voxels with few or even no galaxies assigned. To mitigate this problem, galaxies are represented as point clouds, with each galaxy depicted as a single point characterized by distinct positions and properties. This representation is then processed through a deep neural network called \texttt{Minkowski-PointNet}, which is a PointNet \citep{qi2016pointnet} implementation in the Minkowski Engine \citep{choy20194d}.

PointNet is a neural network architecture that captures the structure of point clouds, a simplified graph with no edges. PointNet is an architecture that can be generalized as DeepSets \citep{DeepSets}, which captures the permutation invariance and equivariance of point clouds \citep{GDL}. Such geometric priors are captured from the 1D convolution layers and the global pooling layers. Despite PointNet's use of rotation and translation invariance to handle point clouds, such procedures are omitted in our approach because of the redshift dependence of features and clustering, as well as the (RA,DEC) dependence of masking. Moreover, to explicitly introduce local properties, we apply the k-nearest-neighbor (kNN) algorithm to survey the characteristics of neighboring galaxies and explicitly add them to the feature vector. Such a step is inevitable since we are not able to perform message-passing between the nodes or the points, as the computational costs involving calculation on the edges are extremely demanding for the mocks comprising more than 150,000 galaxies. Therefore, we add the local information to the feature vector, to enrich the information fed to the machine.\footnote{In contrast to PointNet$++$ \citep{point++}, which uses kNN for grouping and non-uniform sampling of points, we do not adopt such set abstraction layers since the absolute number of galaxies comprising each realization needs to be informed to the machine.}

The Minkowski Engine is a library that efficiently handles sparse tensors, including operations such as auto-differentiation and convolution. Galaxies are grouped and quantized into sparse tensors based on their (RA, DEC, $z$) positions using the engine, where $z$ denotes the redshift. The main advantage of this implementation lies in its ability to handle a variable number of points as inputs to the machine, whereas the original implementation of PointNet operates on fixed sizes. Additionally, it efficiently utilizes memory by grouping galaxies into sparse tensors. 
This approach results in approximately 25\% of the quantized cells containing more than one galaxy, and around 5\% containing more than two galaxies. This strategy effectively preserves the local structure while ensuring better memory consumption and performance.

The specific network layout is illustrated in Figure \ref{fig:NetworkLayout}. The \texttt{Minkowski-PointNet} is capable of receiving point clouds of arbitrary size. The input catalog is transformed into a sparse tensor and passes through a total of five linear layers. Each linear layer is followed by a batch normalization layer \citep{BatchNorm} and a leaky ReLU activation function. The tensor is then passed through the global sum, average, and max-pooling layers and concatenated to a 1536-dimensional vector. Global aggregators are crucial to reflecting the permutation invariance of the neural network. Unlike the original implementation of PointNet, solely using the global max-pooling as the aggregator, we add other aggregators to better capture the embedded information as suggested in \cite{PNA_Graph}. After four consecutive linear layers, the machine predicts the $\Omega_{\rm m}$, $\sigma_{8}$, and their standard deviations, which will be used for implicit likelihood inference.

During the training process, we use the ADAM optimizer \citep{ADAM} with a learning rate of 10$^{-7}$ and a ReduceLRonPlateau scheduler, which reduces the learning rate when the validation loss is not decreased, for a total of 20 epochs. We make use of 80\% of the samples as a training data set and 10\% each as validation and test data sets. We adopt the loss function for implicit likelihood inference as described in \cite{Jeffrey2020}, which is the sum of the following two loss functions, where $y$ is the label and $\sigma^{2}$ the variance.
    \begin{eqnarray}\label{eqn:lfi1}
    L_{1} = \ln \left[\sum_{i \in batch} (y_{i,pred}-y_{i,true})^{2}\right] 
    \end{eqnarray}
    \begin{eqnarray}\label{eqn:lfi2}
    L_{2}=\ln \left[\sum_{i \in batch} ((y_{i,pred}-y_{i,true})^{2}-\sigma_{i}^{2})^{2}\right]
    \end{eqnarray}
By minimizing the combined loss function $L_{\rm{vanilla}}{=}L_{1}{+}L_{2}$, we optimize both prediction accuracy and enable the representation of the second moment, which corresponds to the standard deviation. Such approaches have recently been utilized in many machine learning projects to estimate the model's error in the absence of likelihoods \citep{Cosmo1Gal, HaloMassGNN}.

    \begin{figure*}[t]
    \centering
    \includegraphics[width=0.83\textwidth]{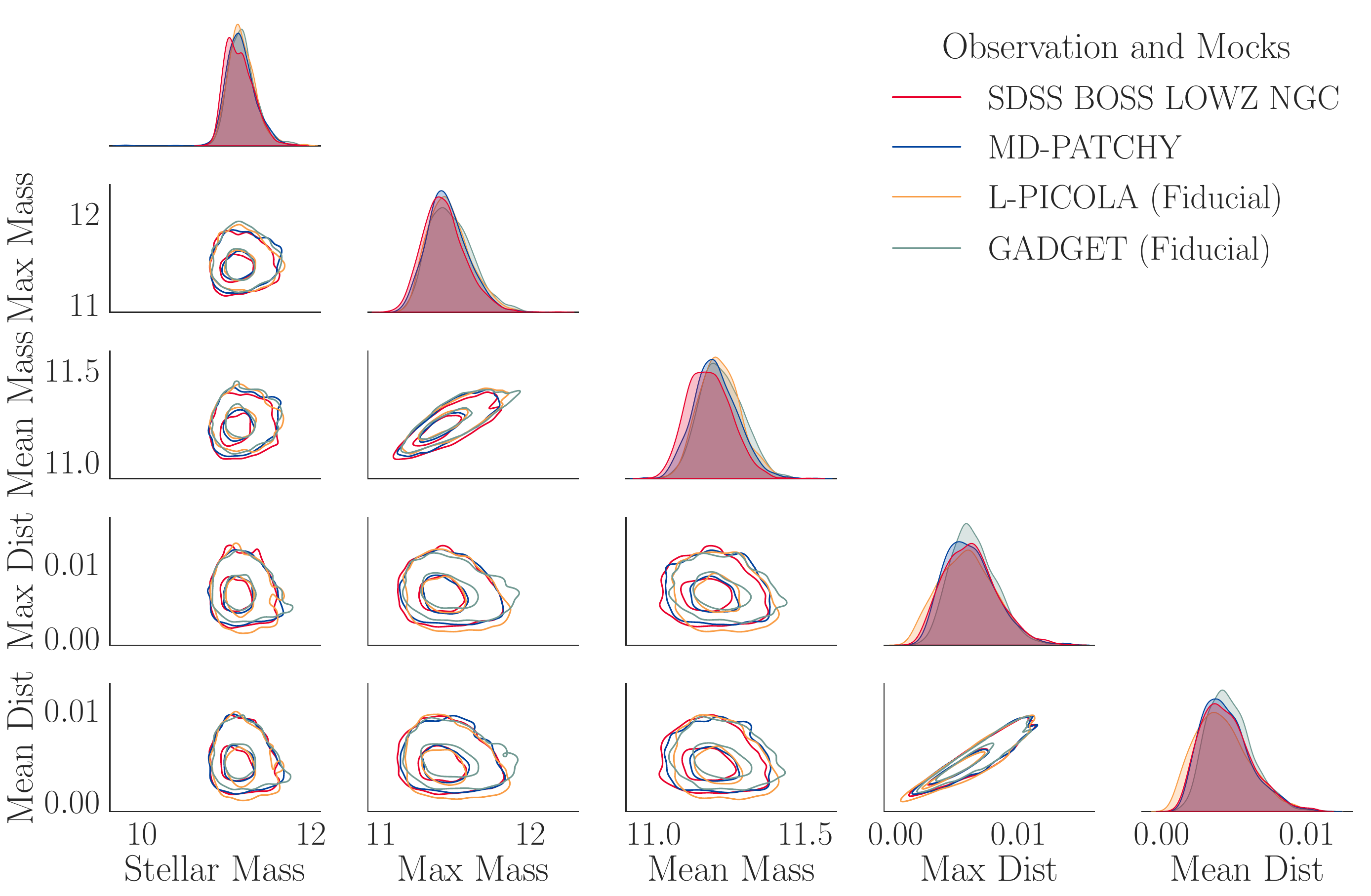}
    \caption{\label{fig:feature_pairs} Pair plot of five features of a galaxy randomly sampled from each mock generated with the \shmr{} model: stellar mass of itself, maximum and mean neighbor masses, and maximum and mean neighbor distances, for a single realization of SDSS BOSS LOWZ NGC (\textit{red}), \mdpatchy{} (\textit{blue}), \texttt{L-PICOLA fiducial} (\textit{orange}), and \texttt{GADGET fiducial} (\textit{green}). 
    The plot shows for 1000 randomly sampled galaxies for each mocks. 
    Masses are in units of $\log(M_{\star}/h^{-1}{\rm M}_{\odot})$ and distances are expressed in terms of the newly assumed metric in redshift space.
    The distribution exhibits fair consistency across the three mocks and the SDSS BOSS LOWZ NGC catalog.
    See sections \ref{sec:backbone} and \ref{sec:features} for more information.}
    \vspace{2mm}
    \end{figure*}

\subsection{Input Features\label{sec:features}}

The input features of galaxies should align with those derivable from observational data. Thus, we utilize the position and stellar mass of each galaxy, as well as information from its neighbors, to extract details about the local environment, following the methodology presented in \cite{MSSM}. Moreover, it is important to note that we do not provide the machine with physical or comoving distances since they already imply a certain cosmology when converted from observed redshifts. Instead, we introduce a transformed position of each galaxy by $(X,Y,Z)=(z\sin(DEC)\cos(RA), z\sin(DEC)\sin(RA), z\cos(DEC))$. The redshift will be re-introduced as one feature, allowing the machine to infer the redshift dependence of features. 

Additionally, we explicitly incorporate information from neighboring galaxies. This addresses the limitations of \texttt{Minkowski-PointNet}, which does not support message-passing between edges due to computational constraints arising from the large number of inputs. By introducing neighboring information, we expect these features to serve as proxies for relational local information. From the nine nearest neighbors, four local features are selected: mean distance, maximum distance, mean stellar mass, and maximum stellar mass. Again, since we apply a metric in the redshift space, the distances become unitless. The redshift and stellar mass of each galaxy are used as point-specific features. In total, the six features are aggregated per galaxy, combining both local and point-specific characteristics. Figure \ref{fig:feature_pairs} displays a pair plot of features with contours for 1000 randomly sampled galaxies for the mocks generated with the \shmr{} model. The distribution exhibits fair consistency across the three mocks and the SDSS BOSS LOWZ NGC catalog. Another comparison between different cosmologies is available in Appendix \ref{sec: diff_cosmo_feat}. Although not displayed for brevity, the \sham{} models exhibit similar levels of consistency in the mocks. 
    
\section{\label{sec:strategies}Training Strategies}

\subsection{Why is Domain Shift Critical?\label{sec:domain_shift}}

The small-scale clustering statistic and the low mass end of a halo mass function may have distortion because of its approximate nature in the \lpicola{} code. This is due to the dispersive behavior of dark matter particles that leads to an imprecise subhalo determination \citep{HowlettMGS}. Moreover, the on-the-fly lightcone simulation restricts us from exploiting the historical information of individual halos. The evolution of individual subhalos can be tracked using merger trees derived from simulation snapshots. From this, accurate modeling of the galaxy-halo connection through SHAM is feasible using $V_{\rm{peak}}$ or $V_{\rm{max}}$, even for dark matter fields generated with {\rm C{\scriptsize OLA}} simulations as opposed to the lightcone simulation
 \citep{Ding2023}. In an attempt to mitigate the intrinsic limitation of the rapid lightcone simulation, \lpicola{}, \cite{Howlett2022} introduce two free parameters to represent the subhalo number and mass ratio. These values are tuned by fitting the power spectrum monopole of the observational catalog. However, since we aim at performing inference rather than fine-tuning simulations to match observational data, such an adaptation step is inapplicable. We can enhance the flexibility of the models by incorporating extra free parameters and marginalizing over them during inference, particularly with the HOD framework. However, this approach restricts the use of stellar mass information used in modeling the stellar mass incompleteness and as features in the neural networks. We plan to address such issues in future work.

\texttt{Minkowski-PointNet} demonstrates strengths in its lack of specific limits on clustering scale, allowing for analysis across a wide range of scales, unlike most studies that impose an upper bound $k_{\rm max}$ \citep{Hahn2022, Hahn2023, Ivanov2020, Philcox2022}. Even CNNs inherently impose an effective clustering scale through voxelization \citep{Lesmos2023}. However, our approach is sensitive to small scales, offering rich clustering information while also being susceptible to small-scale distortions specific to each domain's codes. Therefore, it is critical to regularize the training of neural networks to acquire domain-agnostic knowledge.

Addressing the domain shift is crucial to ensuring the robustness of machines and their applicability to real-world observations. We adapt the machines using prepared suites of mocks: 9000 \lpicola{} mocks as the source, along with either 18 \gadget{} mocks or 2048 \mdpatchy{} mocks as targets. By training them with specific strategies aimed at achieving domain adaptation and generalization, we expect the machines to learn domain-agnostic information. Consequently, they will be capable of extracting representations that can be generalized to multiple domains, particularly observational data.

\subsection{Training Objective: Domain Generalization}

The primary goal of this research is to conduct simulation-based inference on actual observational data using machines robust across different codes for generating mocks. A critical question arises: Can we establish a unified approach to forward modeling our universe and making fair inferences on the cosmological parameters? Unfortunately, current neural networks show apparent discrepancies when applied to other domains \citep{CAMELS-ASTRID28, Shao2023}. However, recent trials in generating domain-adaptive graph neural networks to incorporate various sources have shown the possibility of achieving a more robust inference \citep{Roncoli2023}. 

In the context of transfer learning, which involves the transfer of knowledge from a set of task to relevant tasks, each of the mock suites can be viewed as $n$ mocks sampled from individual domains $\mathcal{D}_{i}$, or $S_{i}=\left\{ (x^{i}_{j}, y^{i}_{j})\right\}^{n}_{j=1} \sim (\mathcal{D}_{i})^{n}$, where $x\in\mathcal{X}, y \in\mathcal{Y}$. $\mathcal{X}$ is the feature space and $\mathcal{Y}$ is the space for labels (cosmological parameters), while $\mathcal{D}_{i} \subset \mathcal{P_{XY}}$ is a joint distribution on $\mathcal{X}$ and $\mathcal{Y}$ \citep{DANN, DGReview}. Our aim is to develop a machine that generalizes across multiple domains, even those unseen during the training phase, particularly the observational catalog. Attempts to test the generalizability of a machine trained on a single domain have been initiated by various projects in astronomy using machine learning and deep learning, referred to as ``robustness tests'' \citep{CAMELS-ASTRID28, Shao2023}. In the language of transfer learning, testing on uninvolved domains in the training phase can be viewed as domain generalization \citep[DG;][]{DGReview}. 

To achieve effective domain generalization, it is crucial that the distributions of the target (unseen domains) and source domains (domains involved in the training phase) are similar, which can be achieved through accurate modeling of mocks and training strategies to extract common features. Due to limitations in the accuracy of \lpicola{} mocks, non-negligible discrepancies exist compared to \gadget{} or \mdpatchy{} mocks. Such domain shift (expressed by $\mathcal{H}$-divergence, $d_{\mathcal{H}}(\cdot,\cdot)$) is crucial in setting the upper bound on the empirical risk of any hypothesis \citep{Ben-David2006, Ben-David2010,Albuquerque2019GeneralizingTU}. Thus, achieving single-domain generalization solely through training on \lpicola{} mocks can be challenging.
  
To enhance the machine's generalization capabilities, we utilize \gadget{} or \mdpatchy{} mocks, which enable the machine to acquire common knowledge. Unlike domain generalization, \gadget{} or \mdpatchy{} mocks are incorporated during the training phase, hence, this approach is termed domain adaptation. By employing a training strategy to learn from the relatively accurate mocks, the neural networks learn consistent semantics from the two domains, and finally generalize on the observational data, unseen at training phase. 

A method includes utilizing the domain-adversarial neural network \citep[DANN; ][]{DANN}, which seeks to derive domain-invariant features through the use of a domain classifier as a regularizer. This technique has recently been adopted for performing classification tasks in the field of astronomy \citep{Huertas2023, Ciprijanovic2020}. However, multiple trials show that DANN still suffers from overfitting and there are discrepancies between domains (see Appendix \ref{sec:dann} for more information). We find that such issues can be effectively mitigated by an alternative training strategy, which will be explained in \my{Section \ref{sec:SA_strategy}}.

    \begin{figure*}[!t]
    \centering 
    \includegraphics[width=1.03\textwidth]{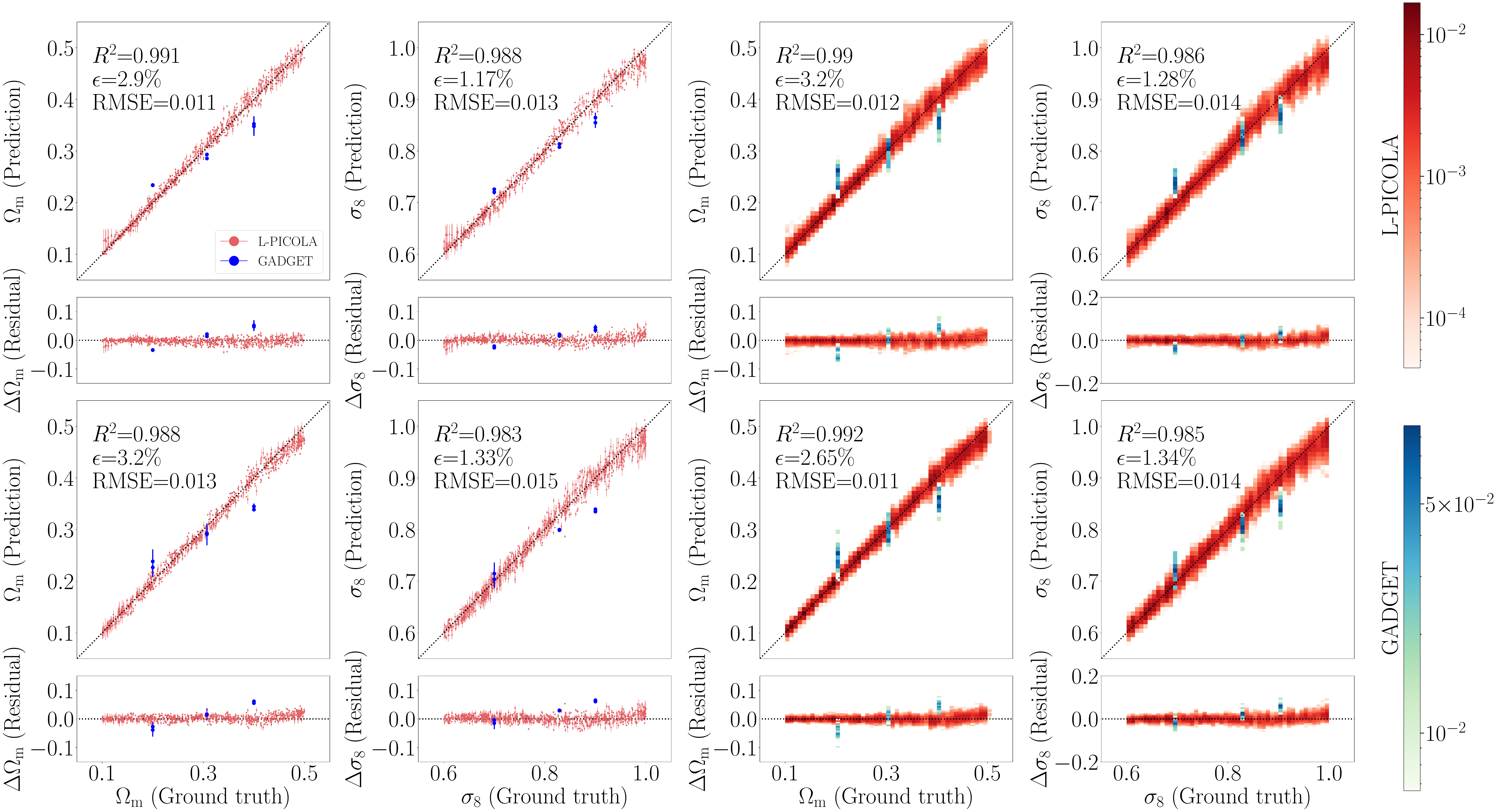}
    \caption{\label{fig:SHMR&SHAM_C+G_SA_accuracy} Comparison of the ground truth and the predicted values of $\Omega_{\rm m}$ and $\sigma_{8}$ on the test set. 
    Predictions are made by \texttt{Minkowski-PointNet} machines with \lpicola{} (\textit{red}) and \gadget{} (\textit{blue}) mocks, trained with the semantic alignment strategy. The \textit{top two panels} display the results from the \sham{} model and the \textit{bottom two panels} display the results from the \shmr{} model.
    The \textit{left two columns} show the results of a single arbitrarily selected machine, while \textit{the right two} show the results for the 25 independently trained machines, with the normalized count expressed in logarithmic color bars. 
    $R^2$, relative error ($\epsilon$) and root-mean-squared-error (RMSE) metrics calculated altogether are shown. 
    Residuals $\Delta y{=}y_{\rm{true}}{-}y_{\rm{pred}}$ are depicted in the \textit{bottom panels}. 
    The machine is trained, validated and tested on the two suites of mocks: \lpicola{} mocks and \gadget{} mocks. 
    Error bars of the \textit{two left columns} indicate the 1$\sigma$ values derived from the implicit likelihood inference. 
    \textit{Black dotted lines} depicts the complete match with null residual.  
    The results from the ensemble of 25 machines for $\Omega_{\rm m}$ and $\sigma_{8}$ show a relative error of $3.20\%$ and $1.28\%$ for the \sham{} model. The \shmr{} model yields $2.65\%$ and $1.34\%$. 
    See Section \ref{sec:perf} for more information.}
    \vspace{3mm}        
    \end{figure*}

\begin{deluxetable*}{llcccc}[t]
\tablewidth{0pt} \label{table:summary}
\tablecaption{Summary of Predictions on the Cosmological Parameters}
\tablehead{\colhead{Models} & \colhead{Training Strategy}& \colhead{$\Omega_{\rm m}$} & \colhead{$\sigma_{8}$} & \colhead{$\epsilon_{\Omega_{\rm m}}$(\%)} & \colhead{$\epsilon_{\sigma_{8}}$(\%)}}
\startdata
\sham{} & Semantic Alignment & $\mathbf{0.339{\pm}0.056}$ & $\mathbf{0.801{\pm}0.061}$& $\mathbf{7.6}$ & $\mathbf{1.2}$ \\
\sham{} & Vanilla  &$0.357{\pm}0.044$ & $0.858{\pm}0.045$& $13.3$ & $5.8$ \\
\shmr{} & Semantic Alignment  &$0.227{\pm}0.035$ & $0.743{\pm}0.039$& $27.9$ & $8.4$\\
\shmr{} & Vanilla  &$0.196{\pm}0.021$ & $0.705{\pm}0.019$& $37.8$ & $13.1$ \\
\cite{Planck2018}   & -   & $0.315{\pm}0.007$ & $0.811{\pm}0.006$ & - & - \\ 
\cite{Ivanov2020}   & -   & $0.295{\pm}0.010$ & $0.721{\pm}0.043$ & - & - \\ 
\enddata
\tablecomments{Summary of cosmological parameter predictions from different models trained with \lpicola{} as \textit{source} and \gadget{} mocks as \textit{target}.
For this work, we refer to the galaxy-halo connection model as the model names, together with the two training strategies: Semantic Alignment (\textit{with domain adaptation}) and Vanilla (\textit{without domain adaptation}). 
The predicted values for each models are given with their respective uncertainties, which include both the uncertainty of individual machine and all 25 independently trained machines combined. 
Together with our main results, we also display the results from the CMB measurements \citep{Planck2018} and the full-shape power spectrum analyses of BOSS \citep{Ivanov2020} for reference. Relative differences $\epsilon_{\Omega_{\rm m}}$ and $\epsilon_{\sigma_{8}}$, calculated with respect to the results of \cite{Planck2018}, are displayed for the models studied in this work.
See Section \ref{sec:SDSS_RESULTS} for more information on the results, and Section \ref{sec:C+G} for the discussion on the comparison between the two training strategies.}
\end{deluxetable*}

\subsection{Training Strategy: Semantic Alignment\label{sec:SA_strategy}}

Our strategy explicitly aligns representations from different domains with similar labels. In other words, given that the samples have similar cosmological parameters, regardless of the selection of simulations, the neural networks extract features that are similar to each other. Aligning the representations can explicitly bring about consistency in terms of their semantics across domains and be effective in domain generalization \citep{Motiian_2017_ICCV}. We adapt the semantic alignment loss in \cite{Motiian_2017_ICCV} to a regression task setup by adding the following loss term: 
\begin{eqnarray}\label{eqn:saloss}
L_{\rm{SA}}{=}\sum_{i{\in} B_{S}}^{}  \sum_{j{\in}B_{T}}^{} \frac{1}{\left\| \mathbf{y}^{\rm{S}}_{i}{-}\mathbf{y}^{T}_{j}\right\|}\left\|g(\mathbf{x}^{S}_{i}){-}g(\mathbf{x}^{T}_{j})\right\| 
\end{eqnarray}
Here, $B_{S}$ and $B_{T}$ represent batches from domains $S$ (source) and $T$ (target), respectively, with $g(\cdot)$ denoting the function that maps input to the representation vector. We apply the semantic alignment loss to the 16-dimensional representation, which can be obtained just before the terminal layer of the neural network, as depicted in Figure \ref{fig:NetworkLayout}.\footnote{In this study, we opted for a reduced representation of 16 dimensions instead of the comprehensive 1536-dimensional representation due to challenges in balancing accuracy and adaptability within our machine learning model. The use of the penultimate layer of linear networks as the representation vector was also used in \citet{Lin2022Bias}. Modifying the architecture of the neural network and performing detailed fine-tuning of hyperparameters are strategies that could enhance adaptability, which we aim to explore in future research.} The generalization strength can be modified by adjusting the weight $\alpha_{p}$ in $L_{\rm{total}}{=}L_{\rm{vanilla}}{+}\alpha_{p}L_{\rm{SA}}$. Here, we slightly modify the adaptation parameter setup proposed by \cite{DANN}, 
\begin{eqnarray}\label{eqn:adaptparam}
\alpha_{p}=\alpha_{0}\left[\frac{2}{1+\exp(-\gamma p)}-1\right]
\end{eqnarray}
where $p$ linearly increases from 0 to 1 as training epochs increase, with $\gamma{=}5$ and $\alpha_{0}{=}5$. This gradual increase in the strength of the adaptation term allows the machine to first gain predictability on the labels before aligning the representations' semantics. Hyperparameters are chosen based on multiple trials to balance the trade-off between prediction accuracy and the strength of domain adaptation. To observe the effectiveness of the alignment process, or domain adaptation, we do not include the samples from the target domain in calculating the vanilla loss (see Equations \ref{eqn:lfi1} and \ref{eqn:lfi2}). Therefore, the labels of targets are only implied to the machine through the semantic alignment loss. When incorporating \gadget{} mocks, we reserve 2/3 of the mocks for training and 1/3 for testing. For \mdpatchy{} mocks, we use 80\% as a training data set and 10\% each for validation and test data sets, same as \lpicola{} mocks.

\section{\label{sec:pred}Prediction of  \texorpdfstring{\tt M\MakeLowercase{inkowski}-P\MakeLowercase{oint}N\MakeLowercase{et}}{Minkowski-PointNet}}
  
In this section, we conduct a series of performance tests of \texttt{Minkowski-PointNet} and make predictions on the cosmological parameters of the observational catalog. Given the stochastic nature of the training outcome arising from the existing trade-off between domain adaptability and the accuracy of individual predictions, we train 25 different machines, whose model parameters are randomly initialized. Before predicting on the actual SDSS BOSS LOWZ NGC data, we perform the same feature sampling by identifying their neighbors, as explained in Section \ref{sec:features}. The designated local and global features are then fed to the trained machines. We compare and discuss the results from a set of machines adapted to different domains, as summarized in Table \ref{table:summary}.
    
\subsection{\label{sec:perf}Performance Tests of  \texorpdfstring{\tt M\MakeLowercase{inkowski}-P\MakeLowercase{oint}N\MakeLowercase{et}}{Minkowski-PointNet}}

Following the training procedures discussed in the previous sections \ref{sec:neuralnets} and \ref{sec:strategies}, machines are trained to predict $\Omega_{\rm m}$, $\sigma_{8}$ and their standard deviations. Figure \ref{fig:SHMR&SHAM_C+G_SA_accuracy} displays test results of machines trained with the semantic alignment strategy on the \lpicola{} and \gadget{} mocks. We present results for an arbitrarily selected single machine and for all 25 individually trained machines. The top two
panels show the results of the \sham{} model, and the bottom two panels show the results of the \shmr{} model. In each case, the upper panels show the comparison between the true and predicted values, while the bottom shows the residual. 
The test results are promising for both $\Omega_{\rm m}$ and $\sigma_{8}$, regardless of the galaxy-halo connection model. The results from the ensemble of 25 machines for $\Omega_{\rm m}$ and $\sigma_{8}$ show a relative error of $3.20\%$ and $1.28\%$ for the \sham{} model and $\epsilon{=}2.65\%$ and $1.34\%$ for the \shmr{} model, respectively. A single machine shows a relative error of $2.90\%$ and $1.17\%$ for the \sham{} model, and $\epsilon{=}3.20\%$ and $1.33\%$ for the \shmr{} model. The difficulty in trying to accurately predict $\sigma_{8}$ seen in recent studies \citep{Cosmo1Gal,villanueva-domingo2022,deSanti2023} is not apparent. 

The blue markers and bins in Figure \ref{fig:SHMR&SHAM_C+G_SA_accuracy} show the domain adaptation results in \gadget{} mocks. Due to semantic loss, we are able to marginalize the selection of domains, which leads to the degradation of accuracy in each simulation set (for more information on the error analysis, see Section \ref{sec:C+G}). Since the machine only implicitly infers the cosmological parameters of the \gadget{} mocks through semantic alignment loss during the training phase, a noticeable bias is observed in the predictions when comparing \gadget{} mocks to \lpicola{} mocks. However, the fact that the machine can make predictions solely by aligning the semantics of the source and target domains is encouraging.

Moreover, considering that the parameter space of input labels is constrained within a range of $\Omega_{\rm m}\in [0.1, 0.5]$ and $\sigma_{8}\in [0.6, 1.0]$, samples like \texttt{GADGET low} and \texttt{high} may encounter asymmetry when calculating the semantic alignment loss. In an extreme scenario, if a sample is characterized by the cosmological parameters $\Omega_{\rm m}=0.6$ and $\sigma_{8}=1.0$, it may suffer from bias due to the lack of samples with larger values of the cosmological parameters. This could lead to center-biased predictions as their representations may experience excessive center-ward pull. Overall, the adaptation results remain quite promising, indicating effective alignment of representations from the two domains by the machine.

    \begin{figure*}[t]
    \centering
    \includegraphics[width=0.93\textwidth]{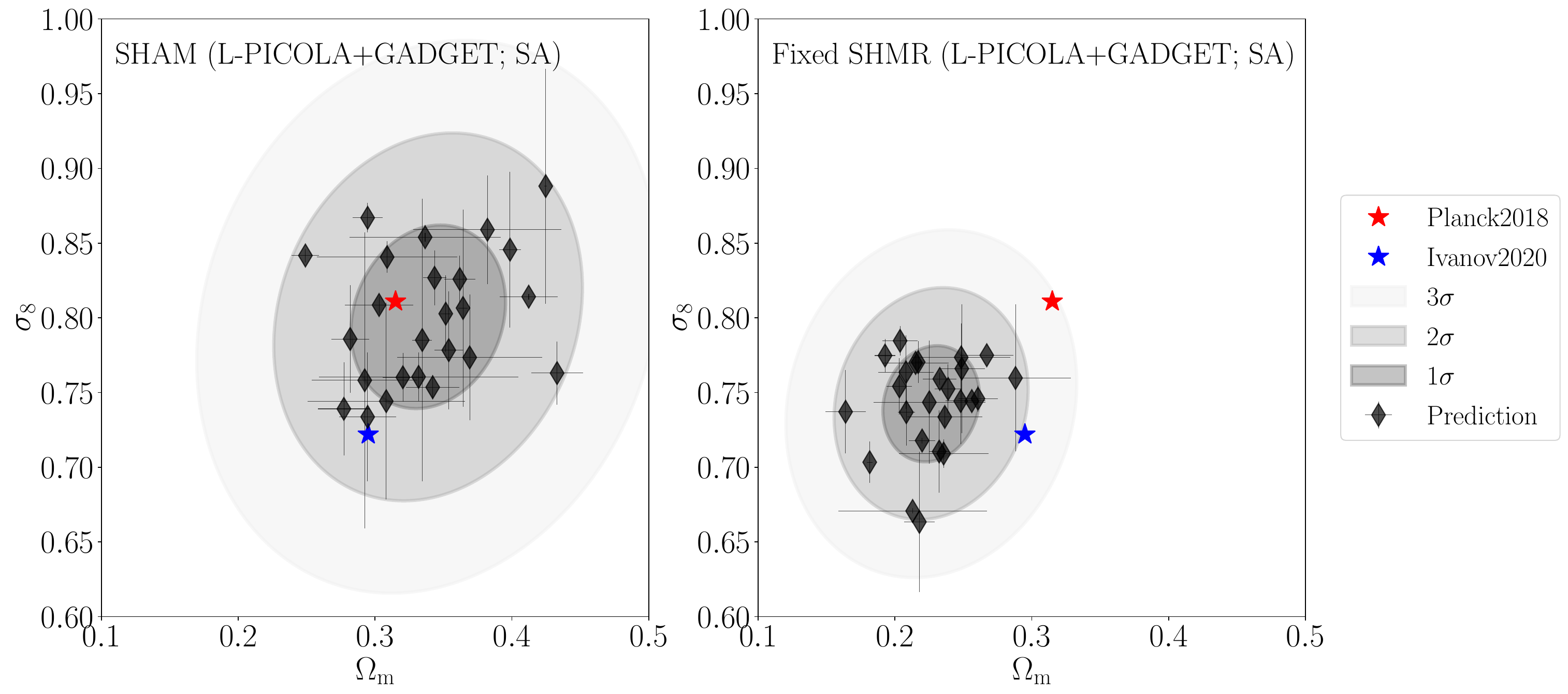}
    \vspace{-1mm}
    \caption{\label{fig:SDSS_FINAL}Prediction on the actual SDSS BOSS LOWZ NGC catalog from the ensemble of 25 independently trained \texttt{Minkowski-PointNet} machines. The \textit{left} figure displays our results when using the \sham{} model, and the \textit{right} figure displays the results when using the \shmr{} model.
    The machines are trained with \lpicola{} and \gadget{} mocks with the semantic alignment (SA) strategy, a domain adaptation and generalization technique that enables the machines to extract consistent features regardless of their simulation domains (see Section \ref{sec:SA_strategy}). 
    Predictions are shown with error bars. 
    A \textit{red star} shows the result from the Planck 2018 \citep{Planck2018} measurements and a \textit{blue star} from \citet{Ivanov2020}. 
    \textit{Elliptic contours} show the bounds of 1$\sigma$, 2$\sigma$, and 3$\sigma$ bounds, calculated from the Gaussian Mixture Model (GMM) to incorporate the individual errors. 
    Our results yield $\Omega_{\rm m}{=}0.339{\pm}0.056$, $\sigma_{8}{=}0.801{\pm}0.061$ (\textit{left, \sham{}}), and $\Omega_{\rm m}{=}0.227{\pm}0.035$, $\sigma_{8}{=}0.743{\pm}0.039$ (\textit{right, \shmr{}}).
    See Section \ref{sec:SDSS_RESULTS} for more information.}
    \vspace{3mm}    
    \end{figure*}

\subsection{Predictions on the SDSS BOSS LOWZ NGC Catalog\label{sec:SDSS_RESULTS}}
In this section, we present predictions on the SDSS BOSS LOWZ NGC Catalog made by the \texttt{Minkowski-PointNet} machines trained with different galaxy-halo connection models and training strategies. Table \ref{table:summary} summarizes the results of the machines trained with \lpicola{} and \gadget{} mocks. Figure \ref{fig:SDSS_FINAL} illustrates the aggregated outcomes of 25 distinct machines, each trained using semantic alignment with \lpicola{} and \gadget{} mocks, alongside benchmark values from \cite{Planck2018} and \cite{Ivanov2020}.\footnote{The main result from \cite{Ivanov2020}, which we cite in Table \ref{table:summary} and figures \ref{fig:SDSS_FINAL}, \ref{fig:SDSS_PATCHY_FINAL}, \ref{fig:SDSS_FINAL_PATCHY}, and \ref{fig:SHMR_DANN}, combines the likelihoods from the Northern Galactic Cap (NGC) and the Southern Galactic Cap (SGC) across two redshift ranges: \textit{low-z} ($z_{\rm eff}=0.38$) and \textit{high-z} ($z_{\rm eff}=0.61$). Although our LOWZ NGC mocks differ from the \textit{low-z} definition, having a lower effective redshift of $z_{\rm eff}=0.29$, the results from the \textit{low-z} NGC used in \citet{Ivanov2020} yield $\Omega_{\rm m}=0.290{\pm}0.017$ and  $\sigma_{8}=0.808{\pm}0.073$ (see sections \ref{sec:SDSS_RESULTS} and \ref{sec:comparison} for more information).} Even within a single training scheme, the predicted results vary significantly between machines, illustrating the stochastic nature of the training process. This suggests that there is degeneracy in the final state of the machine, with multiple configurations exhibiting similar, suboptimal performance. In other words, although different machines demonstrate consistent accuracy and precision on the test set, their predictions on the observational catalog unseen during training phase shows notable variability. This justifies our approach of training multiple machines instead of selecting only those with the best performance.

Next, we compare how the machine predicts on the observational data when trained \textit{with} the domain-adaptive training strategy (semantic alignment) and when trained \textit{without} it (vanilla). For the \shmr{} model, the prediction of the ensemble of 25 machines yield $\Omega_{\rm m}{=}0.196{\pm}0.021$ and $\sigma_{8}{=}0.705{\pm}0.019$ in vanilla scheme, while after applying the semantic alignment loss, $\Omega_{\rm m}{=}0.227{\pm}0.035$ and $\sigma_{8}{=}0.743{\pm}0.039$. The \sham{} model yields $\Omega_{\rm m}{=}0.357{\pm}0.044$ and $\sigma_{8}{=}0.858{\pm}0.045$ in the vanilla scheme, and $\Omega_{\rm m}{=}0.339{\pm}0.056$ and $\sigma_{8}{=}0.801{\pm}0.061$ with semantic alignment. The semantic alignment worsens the precision compared to when not applied, despite increasing the accuracy of prediction, assuming Planck 2018 cosmology as the ground truth. Thus, although the same data sets are being used, the differences in how they are employed to train the machines severely affect the accuracy and precision of prediction on unseen domains. 

The predictions vary significantly depending on the galaxy-halo connection model used to generate the mock catalogs.  Especially, \shmr{} models exhibit considerable divergence from the Planck 2018 cosmology ($\Omega_{\rm m}{=}0.315{\pm}0.007$ and $\sigma_{8}{=}0.811{\pm}0.006$), while \sham{} models are largely in agreement, within the 1$\sigma$ error. Moreover, the $\Omega_{\rm m}$ predicted by the \sham{} models show consistent values with the most recent dark energy survey, \citet{DES2024}, which yields $\Omega_{\rm m}{=}0.352{\pm}0.017$ for the flat $\Lambda$CDM model, a higher value than the Planck 2018 cosmology. Although \sham{} is the most favorable in terms of both accuracy and the absence of any cosmological priors involved in the forward modeling processes, \shmr{} exhibits better precision. This discrepancy likely stems from the additional cosmological priors incorporated via stellar masses in \shmr{} models, as opposed to \sham{} models, which rely solely on clustering information.

This discrepancy can be due to several factors, although the precise cause of this bias in the \shmr{} model remains unclear. One potential reason is that, for the \shmr{} model, regardless of cosmology, any halo with a similar mass will be assigned a similar stellar mass following the SHMR. As discussed in Section \ref{sec:fixed_shmr}, the SHMR from \cite{Girelli2020} was obtained from a different survey, COSMOS, which could also explain the variations. Additionally, the stellar masses of the galaxies in the observational catalog are determined on the basis of the Kroupa IMF \citep{Kroupa2001} with passive evolution from \cite{Maraston2013}, whereas the SHMR we utilized is based on the SMF adjusted for the Chabrier IMF \citep{Chabrier2003} and the stellar population synthesis models from \cite{BruzualandCharlot2003}, which can result in such differences. The exact cause of this discrepancy still being unclear, we stress the limitations of our naive assumption in the \shmr{} model, and that results may vary depending on the galaxy-halo-connection models. Here, we aim to demonstrate the feasibility of inferring without using summary statistics and leave further investigation into the impact of galaxy-halo connection models for future studies.

As mentioned above, when calculating the uncertainty of the inferred parameters, we adopt the most conservative approach. We consider both the error of individual predictions and the 25 independently trained machines, without cherry-picking. However, selecting a single machine that best adapts to and predicts on \gadget{} mocks, characterized by the smallest distance measured by $\sqrt{\Delta\Omega_{\rm m}^{2}+\Delta\sigma_{8}^{2}}$, yields results of $\Omega_{\rm m}{=}0.267{\pm}0.020$ and $\sigma_{8}{=}0.775{\pm}0.0003$ for \shmr{} and $\Omega_{\rm m}{=}0.282{\pm}0.014$ and $\sigma_{8}{=}0.786{\pm}0.036$ for \sham{}. This suggests further potential for performing more precise inference on the cosmological parameters, achieved through the convergence of individual machines and enhanced robustness (see Section \ref{sec:robustness} for a discussion).

    \begin{figure*}[t]
    \centering 
    \includegraphics[width=\textwidth]{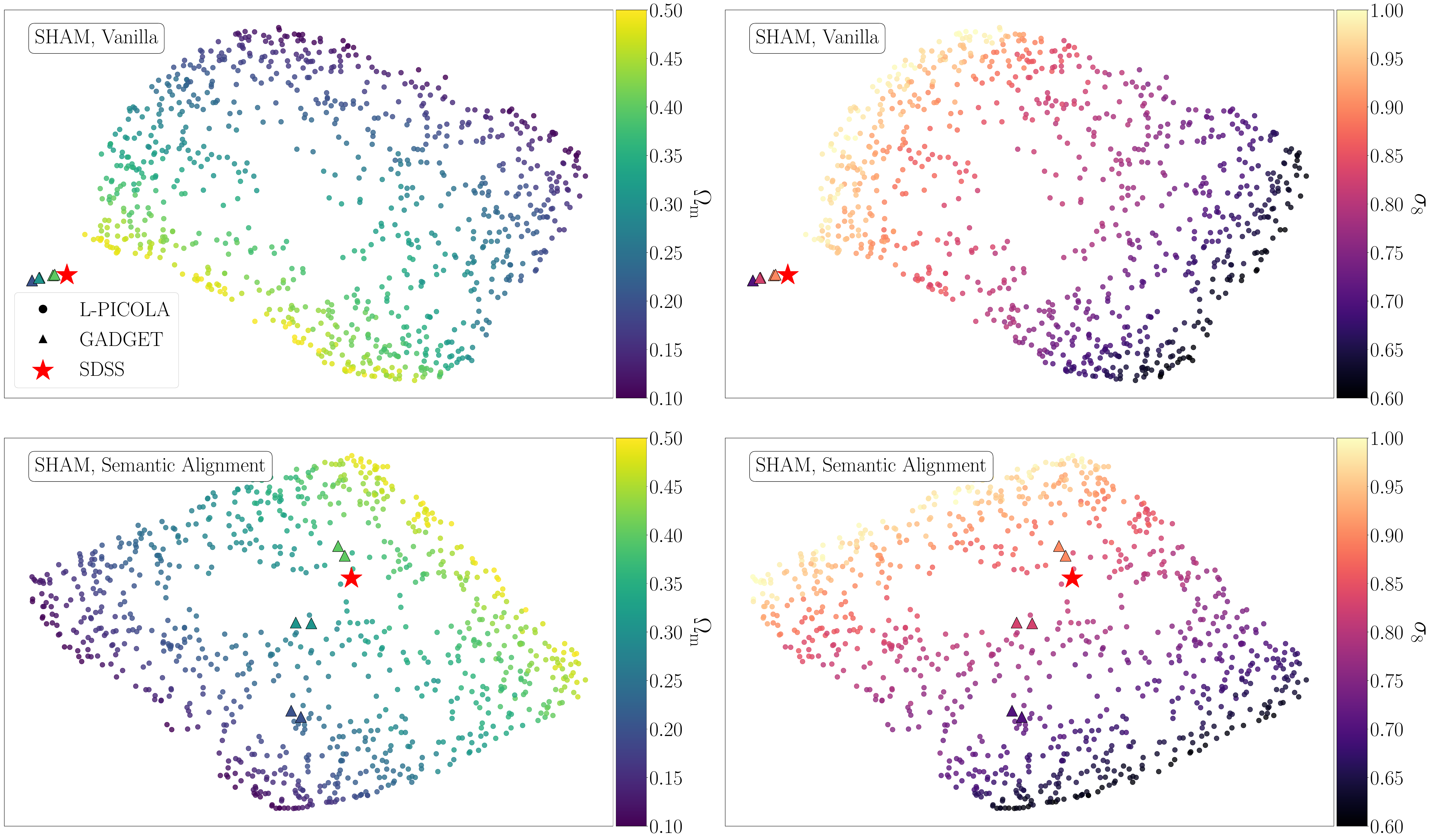}
    \caption{\label{fig:tsne_comparison} A visualization of the latent space configuration from a typical neural network trained with \sham{} \lpicola{} and \gadget{} mocks with the vanilla scheme (\textit{upper panels}) and the semantic alignment strategy (\textit{lower panels}). 
    The 16-dimensional vectors are reduced to two dimensions using the t-SNE algorithm \citep{tsne}. \lpicola{} (\textit{circle}) and \gadget{} (\textit{triangle}) samples are colored according to their cosmological parameters: $\Omega_{\rm m}$ (\textit{left}) and $\sigma_{8}$ (\textit{right}), alongside with SDSS BOSS LOWZ NGC (\textit{red star}). 
    In the \textit{lower panels}, where the semantic alignment strategy is applied, two distinct axes are evident. Along one axis, the parameters gradually change in one direction, while remaining almost independent along the other. This pattern indicates that the two cosmological parameters are effectively represented.
    Moreover, the \gadget{} samples are effectively integrated and generalized in these two panels, in stark contrast to the \textit{upper panels} of the vanilla scheme which show apparent distinction in the distribution.
    See Section \ref{sec:C+G} for more information.}
    \vspace{3mm}    
    \end{figure*}

    \begin{figure*}[t]
    \centering 
    \includegraphics[width=0.88\textwidth]{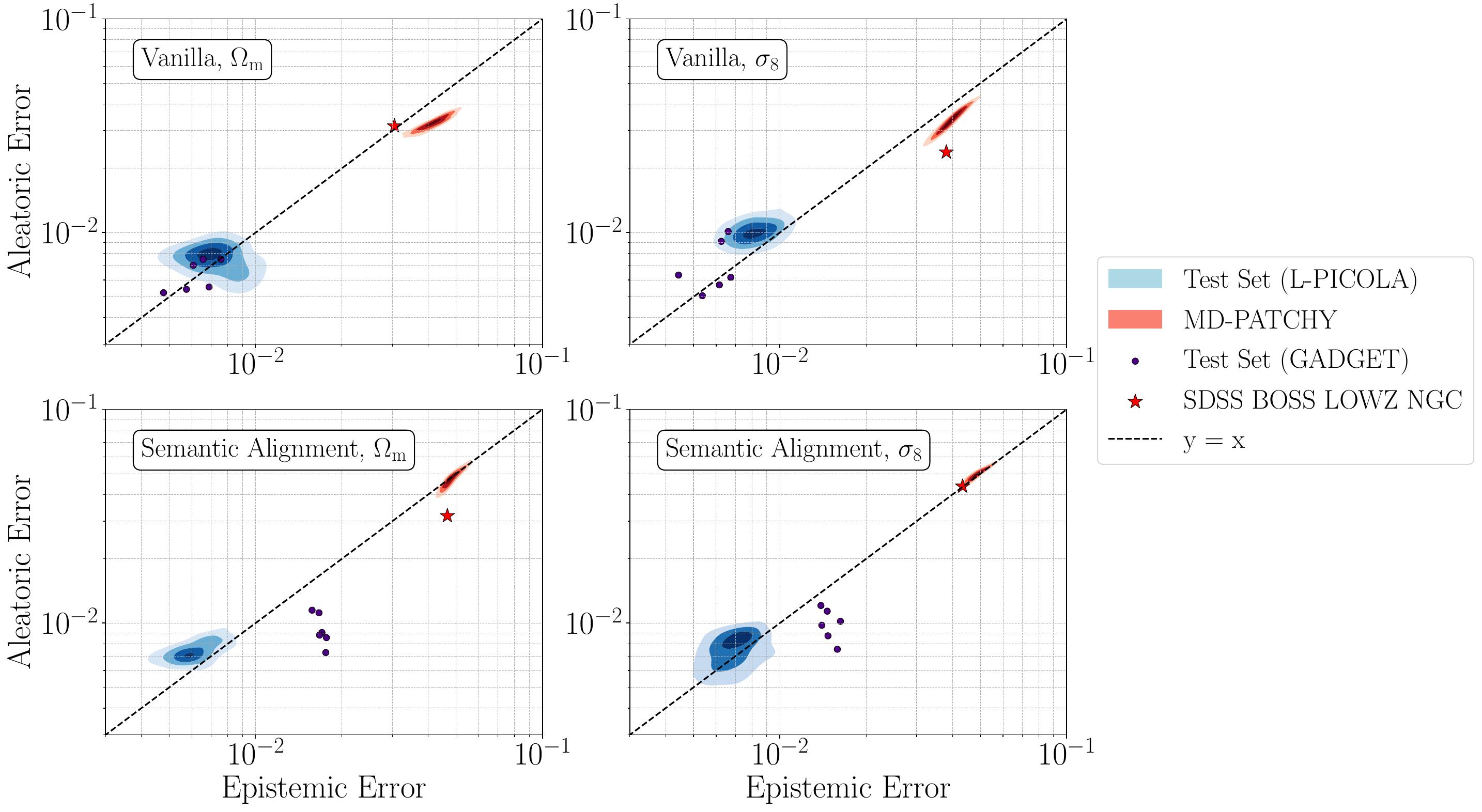}
    \caption{\label{fig:ErrorAnalysis} Comparison of \textit{epistemic} and \textit{aleatoric} error in logarithmic scale, from the ensemble prediction of 25 machines, trained with the vanilla scheme (\textit{upper panels}) and the semantic alignment strategy (\textit{lower panels}). The machines are trained in the two mock suites, \lpicola{} +\gadget{}, with the \sham{} model. The \textit{left two columns} are the results for $\Omega_{\rm m}$ and the \textit{the right two} are the results for $\sigma_{8}$. \textit{Blue contours} represent the test set samples of \lpicola{}, \textit{red contours} the \mdpatchy{} mocks, \textit{dark purple circles} the test set samples of \gadget{}, and \textit{red stars} the SDSS BOSS LOWZ NGC catalog. \textit{Black dotted lines} depict the complete match between the two types of errors. The epistemic errors are calculated by the standard deviation of the predictions on a single input data from the ensemble of 25 machines. The aleatoric errors are calculated by the root mean square of the predicted errors from individual machines. See Section \ref{sec:C+G} for more information.}
    \end{figure*}

\section{Discussion}
\subsection{Effect of Aligning Representations\label{sec:C+G}}

The improvement in generalizability can be attributed to the distribution of different domains aligned in the feature space. To compare the extracted features from machines trained by the vanilla scheme and the semantic alignment strategy, we visually inspect the distributions of their representations in a lower dimension (Jo et al. in prep). Figure \ref{fig:tsne_comparison} exhibits the latent space configuration of the targeted 16-dimensional vector reduced to two dimensions, deduced by the t-distributed Stochastic Neighbor Embedding algorithm \citep[t-SNE;][]{tsne}. In the semantic alignment strategy, the samples are evenly distributed in the reduced dimensions and the parameters gradually change along one direction, while being almost independent in the other direction\footnote{The gaps in the latent space can arise for several reasons. Firstly, the randomness in sampling the parameter space disrupts the dataset's uniformity. Secondly, the dimension reduction technique relies on the distribution's local structure and is inherently nonlinear. Furthermore, because of the discriminative nature of our neural networks, the distribution is not required to be uniform. Generative models such as normalizing flows and variational autoencoders are better suited for accurately modeling the distributions within specific probability distribution functions.}. This behavior naturally suggests that the machine is extracting features and representing them effectively in a way that removes degeneracy and gains predictability in the two parameters.

The vanilla scheme fails to achieve an adaptation of the \gadget{} mocks to the \lpicola{} mocks, resulting in a clear separation between the distributions. The proximity of the observation target to the \gadget{} mocks in comparison to the \lpicola{} mocks demonstrates that the \gadget{} mocks provide a more precise representation of our real universe for the \sham{} model. On the other hand, when the semantic alignment strategy is employed, the two distinct domains blend into a single distribution. Consequently, this supports the claim that the machine is extracting common features from the two domains and less weighting on the domain-specific information, which improves prediction accuracy on the observational data. 

However, there exists a clear trade-off as the semantic alignment loss degrades precision although showing better accuracy. To analyze the effect of semantic alignment on precision, we can first decompose the error into two sources: the \textit{aleatoric} (statistical) error and the \textit{epistemic} (model or systematic) error. The two distinct sources of errors can easily be seen in Figure \ref{fig:SDSS_FINAL}---the aleatoric error estimated from the individual error bars of the machines and the epistemic error from the variance in the prediction from the ensemble of machines.

Figure \ref{fig:ErrorAnalysis} shows the two sources of error for the test sets of \gadget{} and \lpicola{} mocks, which are the domains seen during training phase, and \mdpatchy{} and SDSS BOSS LOWZ NGC samples, unseen during training phase, for the \sham{} model. As we have applied the trained machines to the SDSS BOSS LOWZ NGC catalog, we make inferences on the \mdpatchy{} mocks for further analysis (See Section \ref{sec:robustness} for more information on the results). The epistemic errors are calculated by the standard deviation of the predictions on a single input data from the ensemble of 25 machines. On the other hand, the aleatoric errors are calculated by the root mean square of the predicted errors (see Equation \ref{eqn:lfi2}) of the individual machines. Largely, the aleatoric and epistemic errors have comparable values for both the \lpicola{} test set and the SDSS BOSS LOWZ NGC catalog. However, the errors for the \gadget{} test set show a larger epistemic error compared to the aleatoric error for the semantic alignment training strategy. 

The alignment scheme has a positive effect in reducing errors when predicting in \lpicola{} samples. In particular, the epistemic and aleatoric errors in $\Omega_{\rm m}$ show improvements by 23\% and 4\% each, respectively, and 17\% and 33\% for $\sigma_{8}$. Conversely for \gadget{} samples, epistemic and aleatoric errors on $\Omega_{\rm m}$  show degradation by 92\% and 38\% each respectively, and 86\% and 34\% for $\sigma_{8}$. Thus, we can interpret that the domain-adapted machines exhibit weaker constraints, mostly due to the model-wise uncertainty on the target domain. In other words, the alignment scheme is unstable and can lead to significant variability in the machine's end-of-training state. This considerable variability in model performance on the target domain after domain adaptation can be attributed to the implicit provision of cosmological parameters to the models via the semantic alignment loss, in contrast to the vanilla models. However, the prediction on the unseen observational target shows no significant inclination towards either of the two sources of error. Specifically, the ratio of epistemic to aleatoric error increases by 21\% for $\Omega_{\rm m}$ and decreases by 23\% for $\sigma_8$ after adaptation. Likewise, for the \mdpatchy{} mocks, which are also unseen during the training phase, both epistemic and aleatoric errors arise, but the focus is on the aleatoric error, thus reducing the ratio of epistemic to aleatoric error.

Seen from the analyses above, domain adaptation with semantic alignment improves the overall generalizability on the unseen domains and precision in the source domain while sacrificing precision in target and unseen domains. Although its detailed impact on the precisions are indeed complex, the improvement on generalizability can be mathematically modeled by the \textit{domain generalization error bound} \citep{Albuquerque2019GeneralizingTU,DGReview}. The upper bound of the domain generalization error can also be decomposed into a few sources. Firstly, the machines have to perform well in each of the source domains individually and jointly. Moreover, the source domains should well depict the unseen domain while reducing the discrepancy between the source domains. The discrepancy between the source domains can be explicitly reduced by the semantic alignment as seen from Figure \ref{fig:tsne_comparison}, while the discrepancy between the source and the unseen domain can be reduced with the addition of accurate mocks.\footnote{Precisely, given multiple sources $\mathcal{D}^{i}_{S}$, we define a convex hull $\Lambda_{S}=\{\bar{\mathcal{D}}|\bar{\mathcal{D}}{=}\sum_{i{=}1}^{N}\pi_{i}\mathcal{D}^{i}_{S},\,\pi\in\Delta_{N-1}\}$ with $\Delta_{N{-}1}$ being a $N{-}1$ dimensional simplex. We can then find an optimal distribution $\mathcal{D}^{*}{=}\sum_{i{=}1}^{N}\pi_{i}^{*}\mathcal{D}^{i}_{S}$ where $\pi^{*}$ minimizes the distance between the optimal distribution $\mathcal{D}^{*}$ and the target unseen distribution $\mathcal{D}_{U}$. Therefore, the domain discrepancy between the optimal distribution $\mathcal{D}^{*}$ and the unseen domain $\mathcal{D}_{U}$ measured by the $\mathcal{H}$-divergence term ($d_{\mathcal{H}}(\mathcal{D}^{*}, \mathcal{D}_{U})$), and the discrepancy between the two domains inside the convex hull ($\rm{sup}_{\mathcal{D}', \mathcal{D}''\in\Lambda_{S}}\,d_{\mathcal{H}}(\mathcal{D}',\mathcal{D}'')$) are the two major sources of error. Refer to \cite{Albuquerque2019GeneralizingTU} and \cite{DGReview} for more information.} The vanilla scheme has increased performance on the target sources by distinguishing between the domains, while semantic alignment aligns the distribution at the expense of degraded performance on the target domains. Thus, while domain adaptation shows a significant advantage in that it enables generalization through the alignment of domains, it still suffers from other trade-offs resulting in variability in the machines' end-of-training state, leading to weaker constraints on the cosmological parameters.

   \begin{figure*}[t]
        \centering
        \includegraphics[width=0.93\textwidth]{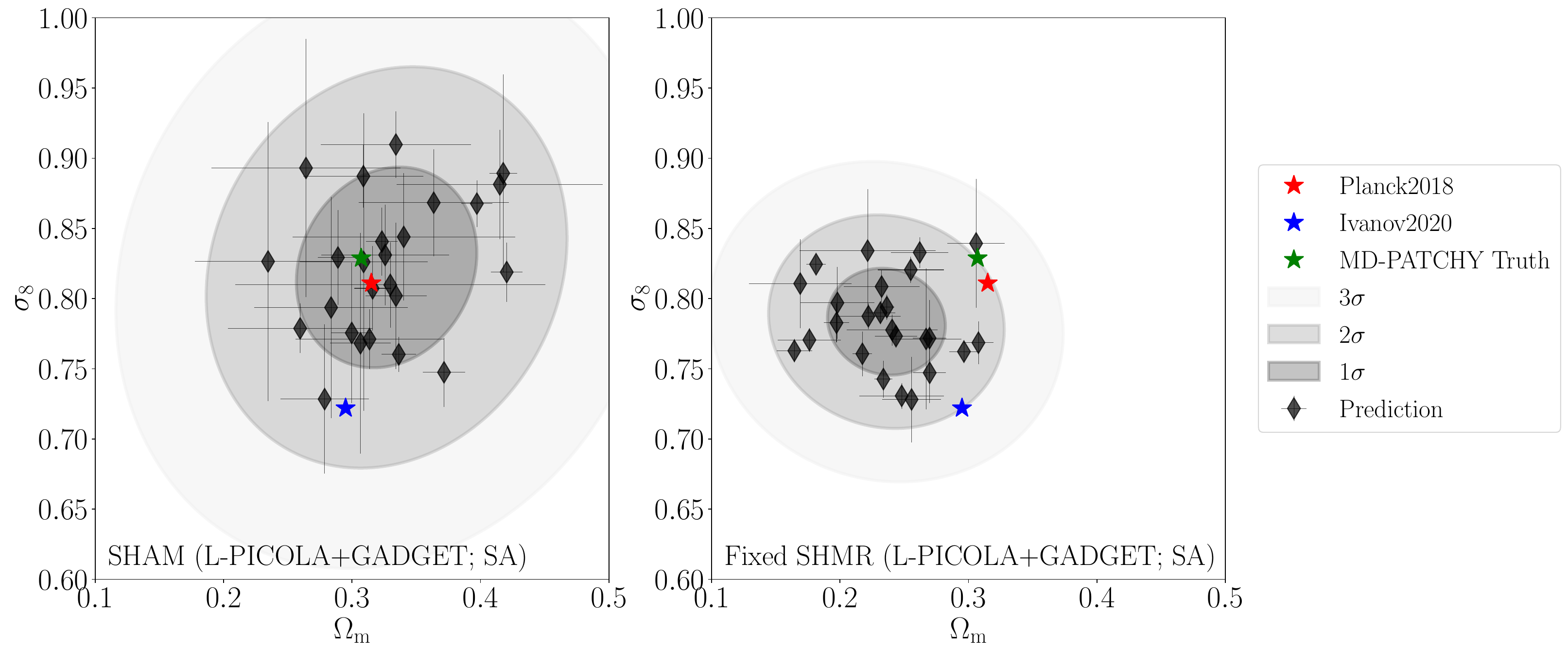}
        \vspace{-1mm}
        \caption{\label{fig:SDSS_PATCHY_FINAL}Prediction on the 2048 \mdpatchy{} mocks from 25 independently trained \texttt{Minkowski-PointNet} machines. The \textit{left} figure displays our results when using the \sham{} model, and the \textit{right} figure displays the results when using the \shmr{} model.
        The machines are trained with \lpicola{} and \gadget{} mocks with the semantic alignment strategy, a domain adaptation and generalization technique that enables the machines to extract consistent features regardless of their simulation domains (see Section \ref{sec:SA_strategy}). 
        Predictions are shown with error bars. 
        A \textit{red star} shows the result from the Planck 2018 \citep{Planck2018} measurements, a \textit{blue star} from \citet{Ivanov2020}, and a \textit{green star} the ground truth values of \mdpatchy{} mocks. Individual error bars include the statistical error attributed to the cosmic variance of the 2048 \mdpatchy{} mocks, which are calculated from the Gaussian Mixture Model (GMM). Individual errors are once again combined by the GMM for the \textit{elliptic contours}, showing the 1$\sigma$, 2$\sigma$, and 3$\sigma$ bounds. 
        Our results yield $\Omega_{\rm m}{=}0.327{\pm}0.070$, $\sigma_{8}{=}0.822{\pm}0.071$ (\textit{left, \sham{}}), and $\Omega_{\rm m}{=}0.236{\pm}0.046$, $\sigma_{8}{=}0.784{\pm}0.038$ (\textit{right, \shmr{}}).
        See Section \ref{sec:robustness} for more information.}
        \vspace{3mm}    
    \end{figure*}

\subsection{Comparison with Previous Studies Using the SDSS BOSS Catalog\label{sec:comparison}}

Our simulation-based inference with neural networks, which replaces the use of summary statistics, yields results that can be compared with several notable studies utilizing the SDSS BOSS catalog. This comparison provides a broader context for evaluating the constraints on cosmological parameters. In the following, we compare our results with previous studies that used summary statistics from the full-shape power spectrum and bispectrum, as well as neural network-based approaches.

Compared to the full-shape power spectrum analyses that yield $\Omega_{\rm m}=0.295{\pm}0.010$ and $\sigma_{8}=0.721{\pm}0.043$ \citep{Ivanov2020} and the bispectrum analyses yielding $\Omega_{\rm m}=0.338^{+0.016}_{-0.017}$ and $\sigma_{8}=0.692^{+0.035}_{-0.041}$ \citep{Philcox2022}, our main results from \sham{} show weaker constraints of $\Omega_{\rm m}{=}0.339{\pm}0.056$ and $\sigma_{8}{=}0.801{\pm}0.061$. However a direct comparison is not possible as our analyses are limited to the BOSS LOWZ NGC sample. In contrast, \citet{Ivanov2020} utilizes the likelihoods combining from the Northern Galactic Cap (NGC) and the Southern Galactic Cap (SGC) across two redshift ranges: \textit{low-z} ($z_{\rm eff}=0.38$) and \textit{high-z} ($z_{\rm eff}=0.61$), and \citet{Philcox2022} both NGC and SGC samples from CMASSLOWZTOT, which combine the LOWZ, LOWZE2, LOWZE3 and CMASS catalogs. 
Although our LOWZ NGC mocks differ from the \textit{low-z} definition, having a lower effective redshift of $z_{\rm eff}=0.29$, the results from the \textit{low-z} NGC used in \citet{Ivanov2020} yield $\Omega_{\rm m}=0.290{\pm}0.017$ and  $\sigma_{8}=0.808{\pm}0.073$.

Next, we compare our results with the recently developed simulation-based inference framework, {\rm S{\scriptsize IM}BIG}, which uses BOSS CMASS samples \citep{Hahn2022, Hahn2023, Lesmos2023}. \citet{Hahn2022} used the power spectrum information up to $k_{\rm max}{=}0.5 h/{\rm Mpc}$ together with normalizing flows, resulting in $\Omega_{\rm m}{=}0.292^{+0.055}_{-0.040}$ and $\sigma_{8}{=}0.812^{+0.067}_{-0.068}$. Compared to these results, we obtain a slightly better constraint on $\sigma_{8}$. On the other hand, \citet{Hahn2023} analyzed the bispectrum monopole up to $k_{\rm max}=0.5 h/{\rm Mpc}$ conducted by using normalizing flows, yielding $\Omega_{\rm m}{=}0.293^{+0.027}_{-0.027}$ and $\sigma_{8}{=}0.783^{+0.040}_{-0.038}$. \my{Therefore, \citep{Hahn2022} and \citep{Hahn2023} explicitly input the cosmological information derived from the clustering statistics at various scales into the machine. In contrast, \citet{Lesmos2023} employ a 3D CNN applied to voxelized galaxy positions in real space, effectively capturing clustering characteristics up to $k_{\rm max}=0.28 h/\rm Mpc$. CNN predictions serve as an intermediate summary statistic, which is then used to generate the final predictions through a flow-based neural network, yielding $\Omega_{\rm m}{=}0.267^{+0.033}_{-0.029}$ and $\sigma_{8}{=}0.762^{+0.036}_{-0.035}$. Our analysis suggests a weaker constraining power compared to previous results.}

However, our study implements a more direct form of simulation-based inference using the embedding extracted by \texttt{Minkowski-PointNet}. As \citet{Lesmos2023} point out, such direct inference from neural network embeddings shows weaker constraints. Thus, recent studies consider the predictions of neural networks as summary statistics and perform additional Bayesian inferences \citep{Gupta2018, Ribili2019, Fluri2019, Lesmos2023}.  Moreover, the major difference in our approach is that we adopt the most conservative form of setting constraints, presenting the ensemble results of 25 individually trained machines instead of a single machine. This highlights the degeneracy of the machines, which show similar performances on known datasets but produce varying predictions on unseen datasets. As mentioned above, using a single machine that is best adapted to the target \gadget{} samples, we obtain comparably tight constraints of $\Omega_{\rm m}{=}0.282 {\pm}0.014$ and $\sigma_{8}{=}0.786{\pm}0.036$.

\subsection{Towards Improved Robustness\label{sec:robustness}}
The ultimate goal of replacing summary statistics with raw input from the mock catalogs for the inference of cosmological parameters would be to give tight and accurate constraints. However, since the neural networks capture the complexities engraved in the input data regardless of the physical importance, such methodology involves advantages and disadvantages at the same time. To maximize the advantage, one must consider building machines robust against the choice of domains. 

An example of robustness is shown in Figure \ref{fig:SDSS_PATCHY_FINAL}, where our domain-adapted machines are applied to the 2048 \mdpatchy{} samples. The results show $\Omega_{\rm m}{=}0.327{\pm}0.070$ and $\sigma_{8}{=}0.822{\pm}0.071$ for the \sham{} model, and $\Omega_{\rm m}{=}0.236{\pm}0.046$ and $\sigma_{8}{=}0.784{\pm}0.038$ for the \shmr{} model. The uncertainties are increased compared to the prediction results on SDSS BOSS LOWZ NGC catalog, partly due to the cosmic variance of the samples. In particular, the predicted values show differences from the SDSS BOSS LOWZ NGC catalog, despite the high degree of similarity of the \mdpatchy{} mocks in the summary statistics. Especially, for the \sham{} model, the machines correctly predict the lower value of $\Omega_{\rm m}$ and the higher value of $\sigma_{8}$ for \mdpatchy{} compared to the observational counterpart assuming Planck 2018 as the ground truth. In contrast to the domain-adapted machines, the vanilla machines yield $\Omega_{\rm m}{=}0.365{\pm}0.055$ and $\sigma_{8}{=}0.875{\pm}0.054$ for the \sham{} model, and $\Omega_{\rm m}{=}0.199{\pm}0.024$ and $\sigma_{8}{=}0.715{\pm}0.016$ for the \shmr{} model. Again, as we have seen from the prediction results on the SDSS BOSS LOWZ NGC catalog, domain adaptation effectively boosts generalizability at the expense of precision. 

To enhance the robustness of neural networks across diverse simulation and observation domains with varying cosmological parameters, we need more samples from the target domains. Currently, insufficient target domain data affects our ability to adapt and generalize effectively, resulting in increased epistemic or model uncertainties, as discussed in Section \ref{sec:C+G}. This in turn leads to degraded precision in the final predictions, as shown in figures \ref{fig:SHMR&SHAM_C+G_SA_accuracy} and \ref{fig:ErrorAnalysis}. Moreover, biases may arise from the discriminative nature of our current neural network model as seen for the \gadget{} samples in Figure \ref{fig:SHMR&SHAM_C+G_SA_accuracy}. Generative models such as normalizing flows and its variants can be helpful in mitigating such biases and better approximate posterior distributions \citep{2022Tang&Ting}. Addressing these biases is crucial to making reliable inferences in data-driven approaches, as emphasized by \cite{Lin2022Bias}.

To accommodate a broader range of cosmological parameters while retaining robustness, not only do we require more sophisticated neural network architectures, but also a focus on the accuracy and correctness of input data. In such data-driven approaches using highly sophisticated neural networks, unreliable input data will distort the extracted domain-agnostic representation. Furthermore, as demonstrated in Section \ref{sec:SDSS_RESULTS}, achieving both precision and accuracy in individual predictions is critical. By improving domain adaptation strategies and utilizing augmented target data, we can potentially enhance the precise inference of cosmological parameters, especially by focusing on reducing the model uncertainties. We plan to explore this potential further in future work.

\subsection{Limatations and Considerations\label{sec:limitation}}
We have demonstrated a proof-of-concept test of inferring cosmological parameters \textit{without} relying on summary statistics, yet there are several limitations and considerations that merit discussion. \lpicola{} mocks, which are our main source domain, show inaccuracies, especially in modeling the halo mass function and small-scale clustering. These inaccuracies are worsened by the simplified assumptions in our galaxy-halo connection models, \sham{} and \shmr{}. Our machine learning models, particularly \texttt{Minkowski-PointNet}, do not enforce explicit cut-offs, making them sensitive to such inaccuracies. Although we introduced \gadget{} mocks and performed domain adaptation to address these issues and improved the models' generalizability, this method involves trade-offs in precision. 

To tackle these challenges, we suggest several strategies. To begin with, enhancing the flexibility of our galaxy-halo connection models by incorporating additional modeling parameters may improve both accuracy and robustness. Secondly, our target domain samples currently lack diversity in the domains and cosmologies, which might limit the generalizability of our models. Addressing this issue involves considering the inclusion of more mock samples from diverse codes, despite the higher computational costs. Additionally, exploring alternative techniques for domain adaptation and generalization could foster improvements in model performance across various datasets. 

\my{Additionally, it is essential to explore the application of our new methodology to a range of galaxy redshift surveys, which vary in observational effects such as color magnitude cuts, survey depths, completeness, and footprints. Given that our mocks are explicitly modeled to include observational effects unique to the SDSS BOSS LOWZ NGC catalog, our present neural network cannot be applied to other observational surveys. In order to enhance the neural network's robustness against varying observations, we could augment our mock dataset with random cuts and masks, along with modifying radial selection functions. We plan to explore these strategies in our upcoming research.}

\section{\label{sec:conc}Summary \& Conclusion}
We propose a novel approach to rapidly model vast quantities of galaxy catalogs through lightcone simulations, while fully incorporating the observational effects of the SDSS BOSS LOWZ NGC catalog and inferring $\Omega_{\rm m}$ and $\sigma_{8}$ from the actual observations using trained neural networks. 
This addresses the question of whether performing simulation-based inference on observed galaxy redshift surveys using neural networks is feasible in the absence of summary statistics, but only with the position and mass information of individual galaxies. 
Our method extends previous works that perform ``robust field-level inference'' on different codes without adopting summary statistics \citep{Shao2023,deSanti2023}, and works that use summary statistics to infer values from the actual galaxy redshift surveys \citep{Hahn2022}.

Using lightcone simulation \lpicola{}, we generate 9000 galaxy catalogs with varying cosmological parameters in a volume of (1.2$h^{-1}$Gpc)$^3$. 
Subhalos are identified using {\rm R{\scriptsize OCKSTAR}}, with each subhalo assumed to host a single galaxy. We propose two models of galaxy-halo connection, \shmr{} and \sham{}. The \shmr{} model assumes a constant star formation efficiency is assumed within a certain halo mass range across different cosmologies, allowing us to identify stellar masses with varying values across different redshift bins. However, the \shmr{} model suffers from the inclusion of cosmological priors, since they are determined from simulations assuming fiducial cosmology. Therefore, we introduce the \sham{} model, free of cosmological priors, which paints the halo catalog by assuming a monotonic relation with the observed catalog. The catalogs undergo further processing to mimic the observational effects of the SDSS BOSS LOWZ NGC catalog, including RSD, survey footprint using the {\rm M{\scriptsize ANGLE}} masks, stellar mass incompleteness (for \shmr{}), radial selection, and fiber collision (Section \ref{sec:sims}).

The results and key takeaways are summarized below. 	\textit{Without} employing summary statistics and using galaxies as point cloud inputs (Section \ref{sec:neuralnets}), we perform implicit likelihood inference \citep{Jeffrey2020} and derive constraints on $\Omega_{\rm m}$ and $\sigma_{8}$ from the SDSS BOSS LOWZ NGC sample. Rapidly generated \lpicola{} mock representations can be aligned with the more accurate \gadget{} mocks to achieve effective domain generalization using the semantic alignment loss (Section \ref{sec:strategies}). Machines trained and adapted independently with \lpicola{} and \gadget{} mocks infer values of $\Omega_{\rm m}{=}0.227{\pm}0.035$ and $\sigma_{8}{=}0.743{\pm}0.039$ for the \shmr{} model and $\Omega_{\rm m}{=}0.339{\pm}0.056$ and $\sigma_{8}{=}0.801{\pm}0.061$ for the \sham{} model, when applied to the SDSS BOSS LOWZ NGC catalog. Despite the divergence in the prediction results from the \shmr{} model, the \sham{} model, which is free of cosmological priors, agrees with the \cite{Planck2018} results within 1$\sigma$ (Section \ref{sec:SDSS_RESULTS} and Figure \ref{fig:SDSS_FINAL}). 

Although the constraints highlighted in Section \ref{sec:limitation} exist, we have demonstrated advancements in performing simulation-based inference on observations \textit{without} the use of any summary statistics. This was primarily achieved by adapting across two different code domains, to extract a unified knowledge applicable to real-world observations. Moving forward, we aim to incorporate precise data from various fields and utilize more advanced models to enhance the robustness of our models. This could potentially establish the new method as a competitive approach in precisely constraining cosmological parameters.

\section*{} 
Jun-Young Lee would like to thank Aleksandra Ćiprijanović, Francisco Villaescusa-Navarro, Yong-uk Cho, Cullan Howlett, Hyeonyong Kim, Seungjae Lee, Jubee Sohn, Jun Yong Park, and Eun-jin Shin for insightful discussions. 
He would also like to thank Francisco-Shu Kitaura and Cheng Zhao for providing the \mdpatchy{} mocks. 
Jun-Young Lee's work was supported by Korea Institute for Advancement of Technology (KIAT) grant funded by the Korean Government (Ministry of Education) (P0025681-G02P22450002201-10054408, Semiconductor–Specialized University).
Ji-hoon Kim’s work was supported by the Global-LAMP Program of the National Research Foundation of Korea (NRF) grant funded by the Ministry of Education (No. RS-2023-00301976).
His work was also supported by the NRF grant funded by the Korea government (MSIT) (No. 2022M3K3A1093827 and No. 2023R1A2C1003244). 
His work was also supported by the National Institute of Supercomputing and Network/Korea Institute of Science and Technology Information with supercomputing resources including technical support, grants KSC-2020-CRE-0219, KSC-2021-CRE-0442 and KSC-2022-CRE-0355.
Jaehyun Lee is supported by the National Research Foundation of Korea (NRF-2021R1C1C2011626).
HSH acknowledges the support by Samsung Electronic Co., Ltd. (Project Number IO220811-01945-01) and by the National Research Foundation of Korea (NRF) grant funded by the Korea government (MSIT), NRF-2021R1A2C1094577.

Funding for SDSS-III has been provided by the Alfred P. Sloan Foundation, the Participating Institutions, the National Science Foundation, and the U.S. Department of Energy Office of Science. The SDSS-III web site is \href{http://www.sdss3.org/}{http://www.sdss3.org/}. SDSS-III is managed by the Astrophysical Research Consortium for the Participating Institutions of the SDSS-III Collaboration including the University of Arizona, the Brazilian Participation Group, Brookhaven National Laboratory, Carnegie Mellon University, University of Florida, the French Participation Group, the German Participation Group, Harvard University, the Instituto de Astrofisica de Canarias, the Michigan State/Notre Dame/JINA Participation Group, Johns Hopkins University, Lawrence Berkeley National Laboratory, Max Planck Institute for Astrophysics, Max Planck Institute for Extraterrestrial Physics, New Mexico State University, New York University, Ohio State University, Pennsylvania State University, University of Portsmouth, Princeton University, the Spanish Participation Group, University of Tokyo, University of Utah, Vanderbilt University, University of Virginia, University of Washington, and Yale University.

\bibliography{sample631}{}
\bibliographystyle{aasjournal}

\appendix

\section{\label{sec: diff_cosmo_feat}Features of realizations with different Cosmological Parameters}

\begin{figure*}[h]
\centering
\includegraphics[width=0.76\textwidth]{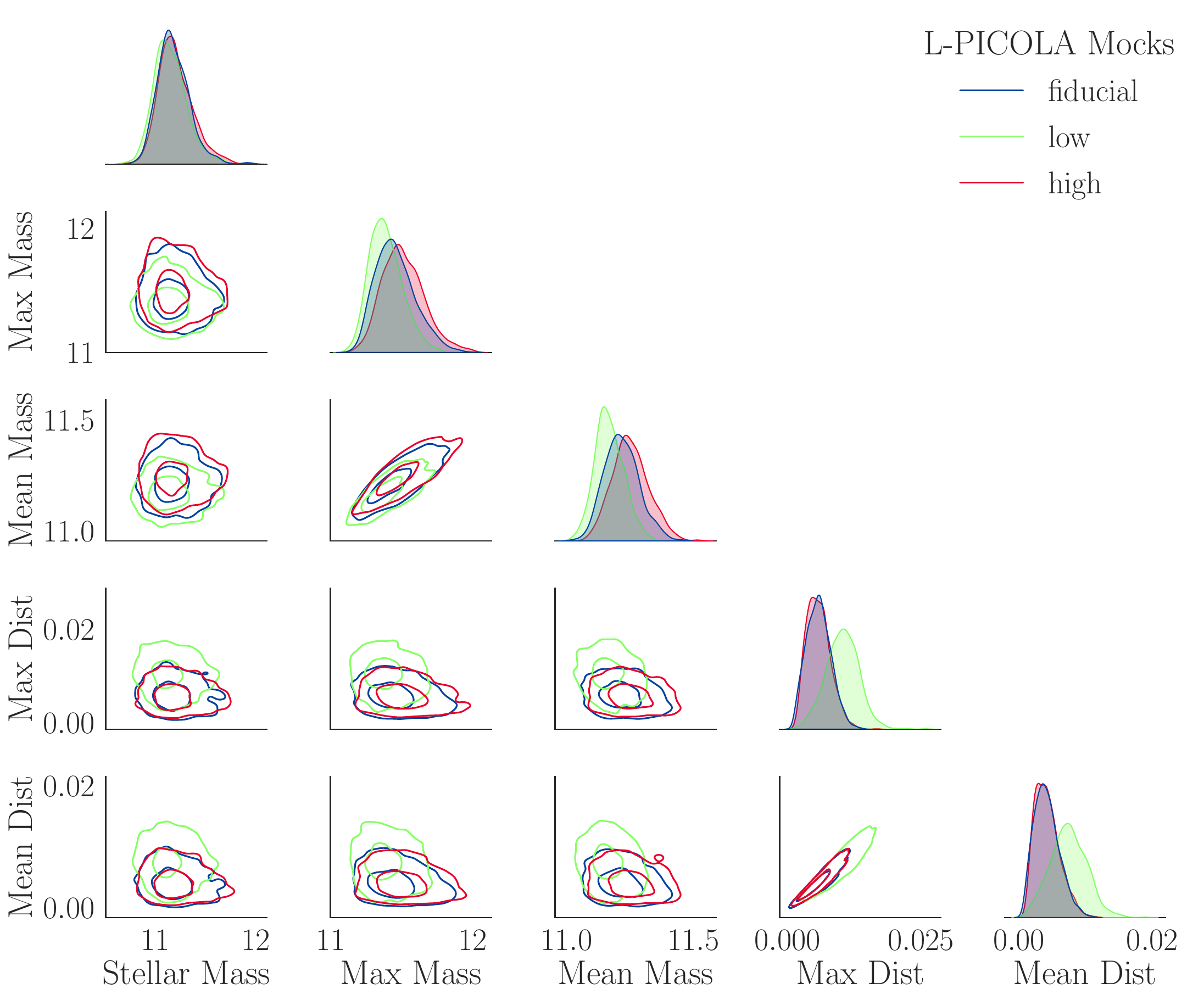}
\caption{\label{fig:othercosmo_feature}Pair plot of five features of a galaxy randomly sampled from each mock generated with the \shmr{} model, similar to Figure \ref{fig:feature_pairs}, but this time for three realizations of different cosmologies named \texttt{high} (\textit{red}; $\Omega_{\rm m}{=}0.4772, \sigma_{8}{=}0.9639$), \texttt{low} (\textit{green}; $\Omega_{\rm m}{=}0.1185, \sigma_{8}{=}0.6163$) and \texttt{fiducial} (\textit{blue}; $\Omega_{\rm m}{=}0.3067, \sigma_{8}{=}0.8238$). 
    The plot shows for 1000 randomly sampled galaxies for each mocks. 
    Masses are in units of $\log(M_{\star}/h^{-1}{\rm M}_{\odot})$ and distances are expressed in terms of the newly assumed metric in redshift space $(X,Y,Z)=(z\sin(DEC)\cos(RA), z\sin(DEC)\sin(RA), z\cos(DEC))$.
    See Appendix \ref{sec: diff_cosmo_feat} for more information.}
\end{figure*}

Features of individual and neighboring galaxies differ across realizations with varying cosmological parameters. In Figure \ref{fig:othercosmo_feature}, we provide the same pair plot as Figure \ref{fig:feature_pairs} but for the \shmr{} model of the \lpicola{} mock suite with different cosmologies: \texttt{high} ($\Omega_{\rm m}{=}0.4772, \sigma_{8}{=}0.9639$), \texttt{low} ($\Omega_{\rm m}{=}0.1185, \sigma_{8}{=}0.6163$), and \texttt{fiducial} ($\Omega_{\rm m}{=}0.3067, \sigma_{8}{=}0.8238$). 
Notice that \texttt{low} deviates the most from \texttt{fiducial}, while \texttt{high} shows a better agreement in all features. 
This tendency becomes most extreme for distances to neighboring galaxies. 
This is due to the deficit of the total number of galaxies for \texttt{low}, which severely affects the separation between the galaxies. Although not displayed for brevity, the \sham{} models exhibit consistency despite differences in the cosmological parameters. Such behavior arises from the fact that, in contrast to the \shmr{} model, the \sham{} model matches the total galaxy count of the mocks to the SDSS BOSS LOWZ NGC catalog.

  \begin{figure*}[!t]
    \centering
    \includegraphics[width=0.90\textwidth]{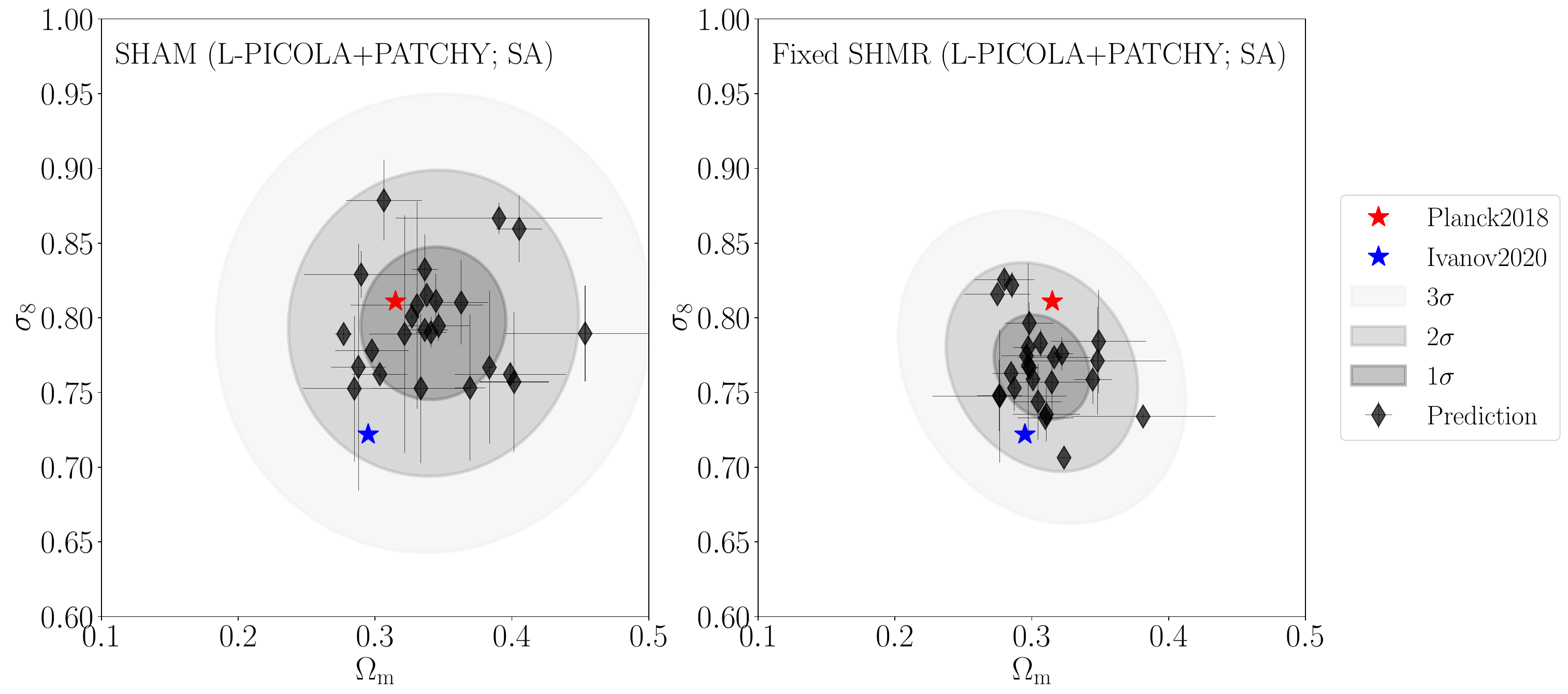}
    \vspace{-1mm}
    \caption{\label{fig:SDSS_FINAL_PATCHY}Prediction on the actual SDSS BOSS LOWZ NGC catalog from the ensemble of 25 independently trained \texttt{Minkowski-PointNet} machines. The \textit{left} figure displays our results when using the \sham{} model, and the \textit{right} figure displays the results when using the \shmr{} model.
    The machines are trained with \lpicola{} and \mdpatchy{} mocks without domain adaptation strategy (Vanilla), a domain adaptation and generalization technique that enables the machines to extract consistent features regardless of their simulation domains (see Section \ref{sec:SA_strategy}). 
    Predictions are shown with error bars. 
    A \textit{red star} shows the result from the Planck 2018 \citep{Planck2018} measurements and a \textit{blue star} from \citet{Ivanov2020}. 
    \textit{Elliptic contours} show the 1$\sigma$, 2$\sigma$, and 3$\sigma$ bounds, calculated from the Gaussian Mixture Model (GMM) to incorporate the individual errors. 
    Our results yield $\Omega_{\rm m}{=}0.343{\pm}0.053$, $\sigma_{8}{=}0.796{\pm}0.051$ (\textit{left, \sham{}}), and $\Omega_{\rm m}{=}0.307{\pm}0.035$, $\sigma_{8}{=}0.767{\pm}0.035$ (\textit{left, \shmr{}}).
    See Section \ref{sec:C+G} for more information.}
    \vspace{3mm}    
    \end{figure*}
    
\section{\label{sec:C+P} Effect of Fine-Tuned  Mocks  \texorpdfstring{\sc MD-\MakeLowercase{patchy}}{MD-patchy}}
We further investigate the possibility of increasing the accuracy and precision via the incorporation of fine-tuned \mdpatchy{} mock samples. Similarly to machines trained with \lpicola{} and \gadget{} mocks, we train 25 different machines using \lpicola{} and \mdpatchy{} with the semantic alignment loss applied. As shown in Figure \ref{fig:SDSS_FINAL_PATCHY}, the results yield $\Omega_{\rm m}{=}0.307{\pm}0.035$ and $\sigma_{8}{=}0.767{\pm}0.035$ for the \shmr{} model, and $\Omega_{\rm m}{=}0.343{\pm}0.053$ and $\sigma_{8}{=}0.796{\pm}0.051$ for the \sham{} model. Compared to when applying the \gadget{} mocks, better precision is achieved for both galaxy-halo connection models. Moreover, especially for the \shmr{} model, the accuracy drastically increases. Indeed, such behavior is well expected, as the machine can learn from the fine-tuned mocks, which better depict the observational sample.

Semantic alignment loss plays an explicit role in reducing the divergence of representations originating from different domains. For example, aligning the representations of $\mathcal{D}_{\rm L-PICOLA}$ and $\mathcal{D}_{\rm MD-PATCHY}$ to be close enough, adding \mdpatchy{} mock samples will have a small impact on increasing the diameter of the convex hull of the domains. Moreover, assuming that the marginal distribution of \mdpatchy{} is relatively similar to the SDSS BOSS LOWZ NGC catalog, the optimal domain, $\mathcal{D}^{*}$, will be weighted towards $\mathcal{D}_{\rm{\mdpatchy{}}}$ and will effectively reduce the generalization risk. Therefore, this confirms not only the importance of aligning the representations from different domains but also the inclusion of accurate mocks involved in the training phase. This effect is maximized for the \shmr{} model, where the initially biased prediction, when trained with the \lpicola{} and \gadget{} domains, significantly alters to produce more accurate results assuming the Planck 2018 cosmology as the ground truth.

However, since \mdpatchy{} mocks are based on a single cosmology, generalization is only effective locally. To train the machines to be globally robust, it is necessary that a multitude of high-fidelity mocks with diverse cosmologies are included as in Section \ref{sec:C+G}. Such inclusion must be made across varying cosmological parameters, unlike the fine-tuned mocks with a single targeted value, as generalization is only performed locally in this case. We leave these aspects of improvement for future work.

\section{Alternative training strategy: Domain Adversarial Training\label{sec:dann}}

An alternative training strategy for domain adaptation and generalization is to extract domain-invariant features through adversarial training. The essence of such a training strategy is to prevent the machine from learning domain-specific information. Here, we employ domain adversarial neural networks \citep[DANN;][]{DANN}, which adds a domain classifier to the backbone of the machine illustrated in Figure \ref{fig:NetworkLayout}. The domain classifier is trained to classify whether the input originates from \lpicola{} mocks or \mdpatchy{} mocks. Moreover, the preceding gradient reversal layer (GRL) enables forward propagation of the domain loss to the feature extractor. Consequently, the feature extractor weights are updated to produce domain-invariant features sufficient to deceive the domain classifier.

    \begin{figure}[h]
    \centering
    \includegraphics[width=0.57\textwidth]{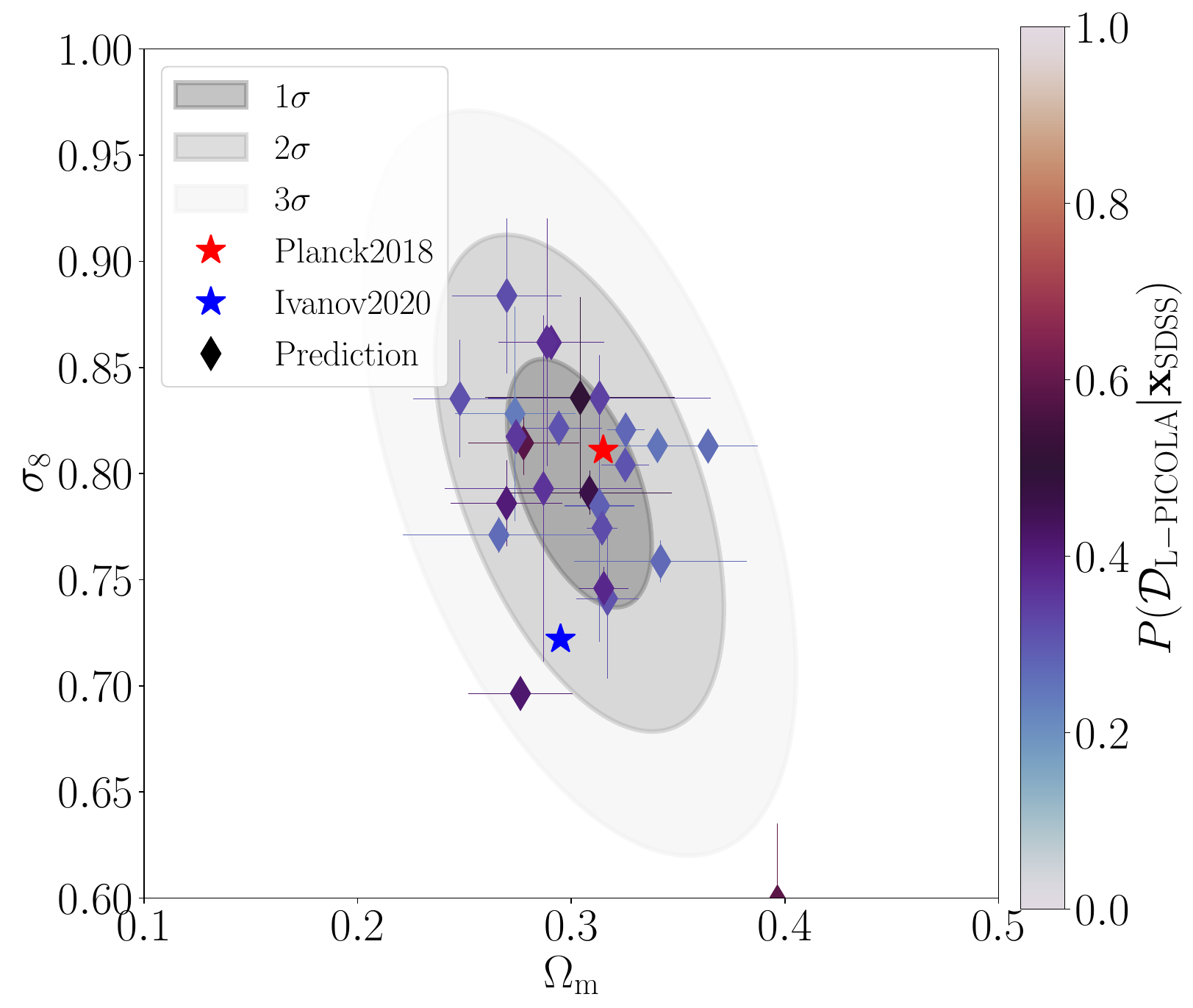}
    \caption{\label{fig:SHMR_DANN}Prediction on the actual SDSS BOSS LOWZ NGC catalog from 25 independently trained \texttt{Minkowski-PointNet} machines, similar to Figure \ref{fig:SDSS_FINAL}, but this time with the domain adversarial neural networks  \citep[DANN;][]{DANN} instead of the semantic alignment strategy. 
    Predictions with error bars are shown and in different colors indicating the probability that the domain classifier classifies as \lpicola{}, $P(\mathcal{D}_{\rm{L-PICOLA}}|\mathbf{x}_{\rm{SDSS}})$. 
    A \textit{red star} shows the result from the Planck 2018 \citep{Planck2018} measurements and a \textit{blue star} from \citet{Ivanov2020}. 
    \textit{Elliptic contours} show the bounds of 1$\sigma$, 2$\sigma$, and 3$\sigma$ bounds. 
    The results yield $\Omega_{\rm m}{=}0.304{\pm}0.033$ and $\sigma_{8}{=}0.795{\pm}0.057$.
    See Appendix \ref{sec:dann} for more information.}
    \end{figure}

In this approach, we leverage the DANN strategy to perform regression tasks in a supervised domain adaptation setup using \lpicola{} and \mdpatchy{} mocks. The loss function of the supervised DANN setup can be mathematically expressed as follows:
\begin{eqnarray}\label{eqn:dann}
L(\theta_{f}, \theta_{r}, \theta_{d}; \mathbf{x}){=}L_{\rm{vanilla}}(G_{r}(\theta_{r}; (G_{f}(\theta_{f}; \mathbf{x}))), \mathbf{y}) \nonumber \\ 
+\alpha L_{\rm{domain}}(G_{d}(\theta_{d}; \mathcal{R}((G_{f}(\theta_{f}; \mathbf{x})))), d)
\end{eqnarray}
where $\theta_{f}$, $\theta_{r}$, $\theta_{d}$ denote the parameters and $G_{f}(\theta_{f},\cdot)$, $G_{r}(\theta_{r},\cdot)$, $G_{d}(\theta_{d},\cdot)$ represent the function of the feature extractor, regressor, and domain classifier. Here, $\mathbf{x}$ represents the input, $\mathbf{y}$ represents the cosmological parameters, and $d$ represents the domain. The GRL $\mathcal{R}(\mathbf{x})$ is a pseudo-function with properties $\mathcal{R}(\mathbf{x}){=}\mathbf{x}$ and $\mathcal{R}'(\mathbf{x}){=} -\mathbf{I}$. Introducing GRL reduces the DANN setup to a single minimization problem.

The terminal layer of the domain classifier passes through a sigmoid activation function, classifying input as \lpicola{} (``1") or \mdpatchy{} (``0") based on a threshold of 0.5. The domain confusion loss $L_{\rm{domain}}$ is calculated using the binary cross-entropy loss with logits, accounting for the imbalance in the size of the data set between each domain. After training, we further train new domain classifiers, each with two trainable layers, for every machine while keeping the weights of the feature extractor frozen. This process allows us to evaluate the classifiability of the extracted features.

Figure \ref{fig:SHMR_DANN} displays the results of the 25 independently trained DANN machines. Individual predictions are colored based on their probabilities as classified by the domain classifier, indicating whether they originated from \lpicola{}, denoted as $P(\mathcal{D}_{\rm{L-PICOLA}}|\mathbf{x}_{\rm{SDSS}})$. The results show $\Omega_{\rm m}{=}0.304{\pm}0.033$ and $\sigma_{8}{=}0.795{\pm}0.057$. However, we observe that compared to the semantic alignment strategy, the distribution between the two domains is not effectively reduced, making it susceptible to overfitting. Consequently, the adequacy of the training scheme can vary depending on the characteristics of the sources and targets and must be used judiciously.



\end{document}